\documentclass[conference,compsoc]{IEEEtran}
\ifCLASSOPTIONcompsoc
  \usepackage[nocompress]{cite}
\else
  \usepackage{cite}
\fi
\ifCLASSINFOpdf
\else
\fi
\usepackage[fleqn]{amsmath}
\usepackage{amssymb}
\usepackage{algorithm}
\usepackage[noend]{algpseudocode}

\usepackage{enumitem}
\newtheorem{theorem}{Theorem}
\newtheorem{lemma}{Lemma}
\usepackage{bm}
\usepackage{multirow}
\usepackage{booktabs}
\usepackage{hyperref}
\usepackage{color}
\usepackage[font=small,labelfont=bf]{caption}
\newtheorem{definition}{Definition}
\newtheorem{corollary}{Corollary}
\newtheorem{proposition}{Proposition}
\usepackage{stmaryrd}

\usepackage{tikz}
\usepackage{tikzpeople}
\usetikzlibrary{decorations.pathreplacing,calligraphy}
\usetikzlibrary{positioning,shadows,arrows,shapes.arrows,shapes.symbols,3d,calc,shapes}
\usepackage{pgfplots} 

\usepackage{ulem}

\begin{document}
%
\title{Bicoptor 2.0: Addressing Challenges in Probabilistic Truncation for Enhanced Privacy-Preserving Machine Learning}

\author{
{\rm Lijing Zhou$^{\dag}$,}
{\rm Qingrui Song$^{\dag}$,}
{\rm Su Zhang$^{\dag}$,}
{\rm Ziyu Wang$^{\dag}$,}
{\rm Xianggui Wang$^\dag$}
{\rm Yong Li$^{\ddagger}$,}\\ 
{\small $\dag$Huawei Technology, China, \{zhoulijing,songqingrui1,zhangsu14,wangziyu13,wangxianggui1\}@huawei.com,}\\
{\small $\ddagger$Huawei Technology Duesseldorf, Germany, \{yong.li1\}@huawei.com}
}
\maketitle

\begin{abstract}
This paper primarily focuses on analyzing the problems and proposing solutions for the probabilistic truncation protocol in existing Private-preserving Machine Learning (PPML) works from the perspectives of accuracy and efficiency. 
  
In terms of accuracy, we reveal that precision selections recommended in some of the existing works are incorrect, which may lead to the inference accuracy being as low as random guessing. We conduct a thorough analysis of their open-source code and find that their errors were mainly due to simplified implementation, more specifically, fixed numbers are used instead of random numbers in probabilistic truncation protocols. Based on this, we provide a detailed theoretical analysis to validate our views. We also propose a solution and a precision selection guideline for future works.
  
Regarding efficiency, we identify limitations in the state-of-the-art comparison protocol, Bicoptor's (S\&P 2023) DReLU protocol, which relies on the probabilistic truncation protocol and is heavily constrained by the security parameter to avoid errors, significantly impacting the protocol's performance. To address these challenges, we introduce the first non-interactive deterministic truncation protocol, replacing the original probabilistic truncation protocol. Additionally, we design a non-interactive modulo switch protocol to enhance the protocol's security. Finally, we provide a guideline to reduce computational and communication overhead by using only a portion of the bits of the input, i.e., the key bits, for DReLU operations based on different model parameters. With the help of key bits, the performance of our DReLU protocol is further improved. We evaluate the performance of our protocols on three GPU servers, and achieve a 10x improvement in DReLU protocol, and a 6x improvement in the ReLU protocol over the state-of-the-art work Piranha-Falcon (USENIX Sec 22). Overall, the performance of our end-to-end (E2E) privacy-preserving machine learning (PPML) inference is improved by 3-4 times.

\end{abstract}

%
\IEEEpeerreviewmaketitle

\section{Introduction} \label{sec:introduction}
The widespread application of PPML is aimed at enhancing privacy protection. Among various PPML techniques, MPC-based PPML is able to disperse data among multiple parties to avoid privacy leakage caused by centralized data collection. However, the inherent characteristic of MPC, which requires interactions among multiple computing parties to complete the computation tasks, makes communication overhead a bottleneck in MPC-based PPML. Existing research has explored various methods to reduce communication overhead, such as using precomputation to improve online performance, designing non-interactive/less-interactive MPC protocols and etc. Nevertheless, these improvements can come with certain drawbacks. Taking truncation protocols~\footnote{Similar to addition and multiplication, truncation is also one of the fundamental building blocks. It is commonly used for precision recovery after fix-point multiplication and also can be applied in the construction of comparison protocols.} as an example, the non-interactive/less-interactive truncation protocols SecureML~\cite{MZ17} and $\text{ABY}^3$~\cite{MR18} demonstrate better performance compared to CryptFlow2~\cite{RRK+-20} and Cheetah~\cite{HLH+-22}. However, these non-interactive/less-interactive truncation protocols may encounter truncation failure due to probabilistic errors. We refer to these types of truncation protocols as ``probabilistic truncation protocols''.The main focus of this paper is to comprehensively analyze and discuss, from the perspectives of accuracy and efficiency, the issues arising from using probabilistic truncation protocols, and we propose corresponding solutions to address these issues. 

\noindent \textbf{The errors in truncation protocols.} 
Existing truncation protocols can be roughly categorized into two types: probabilistic ($\mathsf{trc}_\mathsf{prob.}$) and deterministic ($\mathsf{trc}_\mathsf{determ.}$). Both types of truncation protocols suffer from a 1-bit error issue, caused by the carry bit generated by the truncated part, which is referred to as $e_0$.
Here is an example of using the probabilistic truncation protocol in SecureML~\cite{MZ17} to truncate the last $k$ bits of the input $x\in\mathbb{Z}_{2^\ell}$ and resulting in $e_0$.~\footnote{$\mathsf{trc}(x,4)$ denotes truncating the last 4 bits of $x$ while preserving the sign. $\mathsf{cut}([x]_0,4)$ denotes cutting the last 4 bits of the share $x_0$.}
\begin{align*}
	& x = 0100\ 1011, R = 1010\ 1010, \ell = 8, k = 4, \\
  & [x]_0 = x + R\ \text{mod}\ 2^8 = 1111\ 0101, \\
  & [x]_1 = -R\ \text{mod}\ 2^8 = 0101\ 0110 \\
	& \mathsf{trc}(x,4) \\
  = & (\mathsf{cut}([x]_0,4)\ \text{mod}\ 2^8 - \mathsf{cut}(-[x]_1,4)\ \text{mod}\ 2^8)\ \text{mod}\ 2^8\\
	= &(0000\ 1111 - 0000\ 1010)\ \text{mod}\ 2^8 = 0000\ 0101
\end{align*}
The expected outcome after truncation is $0000\ 0100$ and the real output is $0000\ 0101$. The occurrence of $e_0$ seems inevitable due to the nature of secret sharing.

In addition to $e_0$, probabilistic truncation also has another error, $e_1$, which can directly cause truncation failure. Here is another example to illustrate the significant deviation caused by $e_1$.
\begin{align*}
	& x = 0100\ 1011, R = 1110\ 0000, \ell = 8, k = 4, \\
  & [x]_0 = x + R\ \text{mod}\ 2^8 = 0010\ 1011,\\
  & [x]_1 = -R\ \text{mod}\ 2^8 = 0010\ 0000 \\
	& \mathsf{trc}(x,4) \\
  = & (\mathsf{cut}([x]_0,4)\ \text{mod}\ 2^8 - \mathsf{cut}(-[x]_1,4)\ \text{mod}\ 2^8)\ \text{mod}\ 2^8\\
  = & (0000\ 0010 - 0000\ 1110)\ \text{mod}\ 2^8 = 1111\ 0100
\end{align*}
The actual result of the truncation is $1111\ 0100$, which is far from the expected result of $0000\ 0100$. 

Many PPML works choose probabilistic truncation due to its non-interactive/less-interactive property, while existing deterministic truncation protocols require additional communication overhead. According to the preceding introduction to the errors, $e_0$ appears to be a very minor error with only 1 bit, and we believe its impact on the accuracy of PPML tasks is negligible. On the other hand, $e_1$ is more severe and complex. Previous studies~\cite{W22,MLS+-20,RTP+-22,WTB+-21,BCP+-20,PS20,CRS20,WWP22,TKT+-21} have not extensively addressed $e_1$, typically controlling its occurrence probability through a security parameter to confine its impact to a small, acceptable range on computation tasks. However, this approach does not fundamentally solve the problem of $e_1$ and may give rise to other problems. For instance, selecting appropriate security parameters requires careful consideration of each computation within the entire task. In complex computations, overlooking certain computation processes might lead to erroneous security parameter choices, resulting in a decrease in the accuracy of the computation task. Additionally, security parameters could become bottlenecks or limiting factors in the performance of certain protocols. Our work unveils the essence of $e_1$, providing a detailed explanation of how it arises and, for the first time, presenting its value, which enables us to devise corresponding solutions to eliminate $e_1$ entirely from certain computation process.

\begin{table*}[t]
  \caption{The comparison between the theoretical communication overhead of our DReLU protocol and that of other related works. The PPML runs in the ring size of $\ell=64$. The input precision used in PPML is $\ell_x=\ell_x^\mathsf{int}+\ell_x^\mathsf{frac}=5+26=31$.}
  \label{tab:compare}
  \centering
    \linespread{1.25}
    \footnotesize
      \begin{tabular*}{17.5cm}{@{\extracolsep{\fill}} p{3.5cm} p{2.5cm} p{4cm} l }
        \toprule
        \textbf{Protocol} & \textbf{Preprocessing} & \textbf{Communication Round} & \textbf{One-pass Dominating Communication Cost} \\
        \hline
        Falcon~\cite{WTB+-21}   & Yes  & $4+\log\ell=4+\log_2 64=10$  & $17\ell=17\cdot64=1,088$ bits  \\
        Edabits~\cite{EGK+-20}  & Yes  & $4+\log\ell=4+\log_2 64=10$  & $4\ell-2=4\cdot64-2=254$ bits  \\
        Bicoptor~\cite{ZWC+-23} & No & $2$            & $\ell_x\cdot\ell = 31\cdot64=1,984$ bits  \\
        \hline
        \multicolumn{4}{c}{One-pass Dominating Communication Cost with Key-Bits optimization $(\ell_x=\ell_x^\mathsf{int}+\ell_x^\mathsf{frac}=5+2)$} \\
        \hline
        Falcon~\cite{WTB+-21}   & Yes  & $4+\log\ell=4+\log_2 64=10$  & $16\cdot(\ell-24)+\ell=16\cdot40+64=704$ bits  \\
        Edabits~\cite{EGK+-20}  & Yes  & $4+\log\ell=4+\log_2 64=10$  & $2\cdot(\ell-24-1)+2\cdot\ell=2\cdot39+128=206$ bits  \\
        Bicoptor~\cite{ZWC+-23} & No & $2$            & $\ell_x\cdot\ell = 7\cdot64=448$ bits  \\
        \textbf{Bicoptor 2.0}   & No & $2$            & $(\ell_x + 1)\cdot(\ell_x + 1)= 8\cdot8=64$ bits  \\
        \bottomrule
	  \end{tabular*}
\end{table*}

\subsection{Related Works}\label{suc:relatedwork}
\noindent \textbf{MPC-based PPML.} The current works on secure multi-party computation in PPML mainly focus on two-party, three-party, and four-party settings. The representative works for two-party setting are SecureML~\cite{MZ17}, Delphi~\cite{JVC18}, Chameleon~\cite{RWT+-18}, GAZZLE~\cite{MLS+-20}, CryptFlow2~\cite{RRK+-20}, ABY2.0~\cite{PSS+-21}, Cheetah~\cite{HLH+-22}, and Li et al.~\cite{LCH+-22}, for three-party setting are SecureNN~\cite{WGC19}, Falcon~\cite{WTB+-21}, $\text{ABY}^3$~\cite{MR18}, ASTRA~\cite{CCP+-19}, BLAZE~\cite{PS20}, and CryptFlow~\cite{KRC+-20}, and for four-party setting are Fantastic~\cite{DEK21}, SWIFT~\cite{KPP+-21}, FLASH~\cite{BCP+-20}, and Trident~\cite{CRS20}.


\noindent \textbf{Truncation in PPML.} The non-interactive probabilistic truncation protocol proposed by SecureML~\cite{MZ17} is designed for 2-party scenarios. $\text{ABY}^3$~\cite{MR18}, on the other hand, proposes an interactive probabilistic truncation protocol that is suitable for multi-party (n-party) scenarios. Additionally, CryptFlow2~\cite{RRK+-20} and Cheetah~\cite{HLH+-22} propose two 2-party deterministic truncation protocols, but they require interaction and the communication overhead is very heavy.

\noindent \textbf{Application of Comparison Protocols in PPML.} Comparison protocols, much like multiplication operators in linear layers, play a crucial role in non-linear layers as one of the fundamental building blocks in PPML. Based on comparison protocols, various widely used fundamental protocols in PPML such as DReLU, ReLU, MAX and etc. can be constructed. Presented below are some relevant state-of-art works. The Falcon protocol is proposed to convert computations in $\mathbb{Z}_{2^\ell}$ to computations in $\mathbb{Z}_p$ for smaller prime $p$, to improve the performance of DReLU. However, Falcon's preprocessing is heavy, requiring large-scale distributed methods to generate multiple preprocessed materials offline. Although Edabits~\cite{EGK+-20} and Rabbit~\cite{MRV+-21} further improve the performance of the online phase of DReLU using different types of preprocessing, they both invoke a binary less than circuit~\cite{bitadder}, resulting in a logarithmic communication round complexity. The design of the DReLU protocol proposed in Bicoptor~\cite{ZWC+-23} differs from conventional protocols, e.g., Edabits~\cite{EGK+-20} or Rabbit~\cite{MRV+-21}, in that it does not invoke a binary less than circuit~\cite{bitadder}. Unlike previous ones, it requires only a constant number of communication rounds and does not involve preprocessing. The upper section of Tab.~\ref{tab:compare} summarizes the one-pass dominating communication overhead~\footnote{The communication quantity we calculated refers to the amount of communication sent by a single party.} of previous works on the DReLU protocol, and in Appendix~\ref{app:calc} we provide a comprehensive explanation of our methodology for calculating these theoretical communication costs.

\subsection{Our Contributions} \label{sub:ourcontribute} 
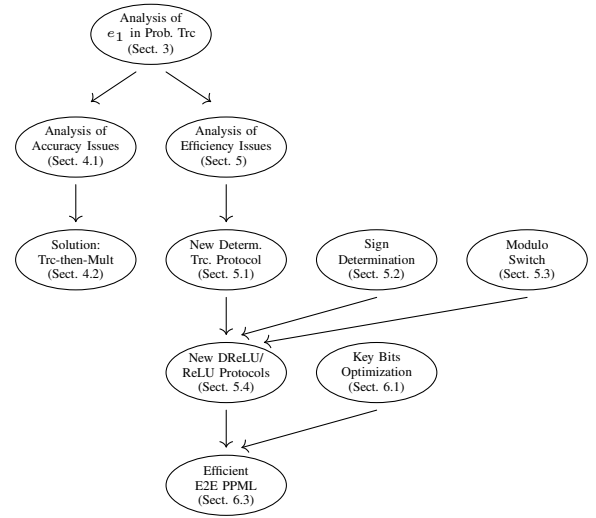
\begin{figure}[ht!]
  \centering
  \begin{tikzpicture}[x=1cm,y=1cm,cap=round,align=center,
      fact/.style={rectangle, draw, rounded corners=1mm, fill=white, drop shadow,
            text centered, anchor=center, text=black},growth parent anchor=center,
      fact2/.style={rectangle, draw, rounded corners=1mm, fill=white,
            text centered, anchor=center, text=black},growth parent anchor=center]
  
  

    \draw (-2,0.5) ellipse (0.8 and 0.4); 
    \node at (-2,0.7) [draw=none, anchor=center]  {\tiny  Analysis of};
    \node at (-2,0.5) [draw=none, anchor=center]    {\tiny $e_1$ in Prob. Trc};
    \node at (-2,0.3) [draw=none, anchor=center] {\tiny (Sect.~\ref{sec:analysis})};
    \draw [<-] (-2.8,-0.4) -- (-2.2,0);
    \draw [<-] (-1.2,-0.4) -- (-1.8,0);

    \draw (-3,-1) ellipse (0.8 and 0.4); 
    \node at (-3,-0.8) [draw=none, anchor=center]  {\tiny  Analysis of};
    \node at (-3,-1) [draw=none, anchor=center]    {\tiny  Accuracy Issues};
    \node at (-3,-1.2) [draw=none, anchor=center]  {\tiny (Sect.~\ref{sub:bug})};
    \draw [<-] (-3,-2) -- (-3,-1.5);
    \draw [<-] (-1,-2) -- (-1,-1.5);

    \draw (-3,-2.5) ellipse (0.8 and 0.4); 
    \node at (-3,-2.3) [draw=none, anchor=center]  {\tiny  Solution:};
    \node at (-3,-2.5) [draw=none, anchor=center]    {\tiny  Trc-then-Mult};
    \node at (-3,-2.7) [draw=none, anchor=center]  {\tiny (Sect.~\ref{sub:linear})};

    \draw (-1,-1) ellipse (0.8 and 0.4); 
    \node at (-1,-0.8) [draw=none, anchor=center]  {\tiny  Analysis of};
    \node at (-1,-1) [draw=none, anchor=center]  {\tiny  Efficiency Issues};
    \node at (-1,-1.2) [draw=none, anchor=center]  {\tiny (Sect.~\ref{sec:drelu}) };
    \draw [<-] (-1,-3.5) -- (-1,-3);
    \draw [<-] (-0.8,-3.5) -- (1,-3);
    \draw [<-] (-0.5,-3.6) -- (3,-3);

    \draw (-1,-2.5) ellipse (0.8 and 0.4); 
    \node at (-1,-2.3) [draw=none, anchor=center]  {\tiny  New Determ.};
    \node at (-1,-2.5) [draw=none, anchor=center]    {\tiny  Trc. Protocol};
    \node at (-1,-2.7) [draw=none, anchor=center]  {\tiny (Sect.~\ref{sub:newtrc})};

    \draw (1,-2.5) ellipse (0.8 and 0.4); 
    \node at (1,-2.3) [draw=none, anchor=center]  {\tiny  Sign};
    \node at (1,-2.5) [draw=none, anchor=center]    {\tiny  Determination};
    \node at (1,-2.7) [draw=none, anchor=center]  {\tiny (Sect.~\ref{sub:cmp})};

    \draw (3,-2.5) ellipse (0.8 and 0.4); 
    \node at (3,-2.3) [draw=none, anchor=center]  {\tiny  Modulo};
    \node at (3,-2.5) [draw=none, anchor=center]    {\tiny  Switch};
    \node at (3,-2.7) [draw=none, anchor=center]  {\tiny (Sect.~\ref{sub:modswitch})};

    \draw (-1,-4) ellipse (0.8 and 0.4); 
    \node at (-1,-3.8) [draw=none, anchor=center]  {\tiny  New DReLU/};
    \node at (-1,-4) [draw=none, anchor=center]    {\tiny  ReLU Protocols};
    \node at (-1,-4.2) [draw=none, anchor=center]  {\tiny (Sect.~\ref{sub:drelu})};
    \draw [<-] (-1,-5) -- (-1,-4.5);
    \draw [<-] (-0.8,-5) -- (1,-4.5);
    
    \draw (1,-4) ellipse (0.8 and 0.4); 
    \node at (1,-3.8) [draw=none, anchor=center]  {\tiny  Key Bits};
    \node at (1,-4) [draw=none, anchor=center]    {\tiny  Optimization};
    \node at (1,-4.2) [draw=none, anchor=center]  {\tiny (Sect.~\ref{sub:keybit})};

    \draw (-1,-5.5) ellipse (0.8 and 0.4); 
    \node at (-1,-5.3) [draw=none, anchor=center]  {\tiny  Efficient};
    \node at (-1,-5.5) [draw=none, anchor=center]    {\tiny  E2E PPML};
    \node at (-1,-5.7) [draw=none, anchor=center]  {\tiny (Sect.~\ref{sub:e2etest})};

  \end{tikzpicture}
  \caption{The Overview of This Paper. ``Prob. Trc.'' stands for Probabilistic Truncation; ``Determ. Trc.'' stands for Deterministic Truncation; ``Trc-then-Mult'' stands for truncate then multiply.}
  \label{fig:paperoverview}
\end{figure}

\begin{itemize} [leftmargin=*]
  \item \textbf{In-depth theoretical analysis of truncation protocol and insights into $e_1$ elimination.} We would like to present our theoretical contributions regarding the truncation protocol. While previous works have introduced the protocol and identified the conditions under which it functions correctly, they lacked a comprehensive analysis of the protocol's underlying principles and design rationale. Through our research, we thoroughly examine the essence of the truncation protocol, enabling us to determine the value of $e_1$ for the first time. This breakthrough discovery leads us to propose innovative methods for eliminating $e_1$, significantly enhancing the protocol's performance and ensuring more accurate and efficient PPML computations. Moreover, such analysis provides future multi-party protocol designers with clearer theoretical functions and directions for their work.

  \item \textbf{Analyzing two probabilistic truncation protocols, exposing the errors in parameter selection of existing works and the reason behind them, and proposing a solution in PPML linear layer.} Firstly, we analyze two truncation protocols proposed by SecureML~\cite{MZ17} and $\text{ABY}^3$~\cite{MR18} from a theoretical perspective and have certain expectations on the inference accuracy for different parameter choices. We point out that many existing works choose inappropriate parameters, but the inference accuracy does not decrease, which is not in line with our expectations. This is because they use fixed numbers for their preprocessed random pair which conceals the truncation error $e_1$. We rerun their experiments using random numbers and the results are consistent with our theoretical analysis (Fig.~\ref{fig:aby3trc} and Tab.~\ref{tab:accuracy}). We provide a solution of truncate-then-multiply instead of multiply-then-truncate for using the original probabilistic truncation protocol while avoiding the impact of the truncation error $e_1$ on linear layers.

  \item \textbf{Proposing a new non-interactive deterministic truncation protocol and a new modulo-switch protocol, and conducting an efficient DReLU protocol based on them.}
  \begin{itemize} [leftmargin=*]
    \item After identifying the cause of errors in SecureML's truncation protocol~\cite{MZ17} through theoretical analysis, we propose a  non-interactive deterministic truncation protocol, which eliminates the truncation error $e_1$, thereby removing the limitation on parameter selection and allows us to choose smaller rings to reduce communication costs. Based on our newly proposed truncation protocol, we construct the Bicoptor 2.0 DReLU protocol, which reduces the communication overhead from $\ell_x\cdot\ell=31\cdot64\approx2,000$ to $31\cdot 31\approx 1,000$ bits.
    \item We introduce a non-interactive modulo-switch protocol which does not require preprocessing to enhance the security and ensure the correctness of our DReLU protocol. Such a protocol has transformed the computations that were originally performed in $\mathbb{Z}_{2^{\ell_x}}$ into computations in $\mathbb{Z}_p$ for prime $p$, where $\log_2p=\ell_x+1$.
    \item Our DReLU protocol is precision-tunable, with the protocol overhead proportional to the data precision. After conducting extensive experiments, we have found that some models can achieve optimal DReLU performance with almost no loss in accuracy when the precision is set to $\ell_x^{\mathsf{int}} + \ell_x^{\mathsf{frac}} = 7$.~\footnote{Note that this precision selection is specific to the four models in our experiments. For other models, the precision selection may differ but the method remains the same.} Such optimization is introduced as the ``key bits optimization''. It can be extended to other works, but enhancements obtained may be limited without our deterministic truncation. For further details on this optimization, please refer to Sect.~\ref{sub:keybit}.
  \end{itemize}
  We also want to emphasize that, like Bicoptor~\cite{ZWC+-23}, our new Bicoptor 2.0 DReLU protocol does not require preprocessing.
  Based on these three contributions, our new DReLU protocol has a theoretical one-pass dominating communication cost of only $(\ell_x+1)\cdot(\ell_x+1)=8\cdot8=64$ bits, and we evaluate our protocol on GPU, which achieves an order of magnitude improvement over P-Falcon~\cite{WWP22,piranha-code}.

  \item \textbf{E2E PPML inference implementation and evaluation on GPU.} We deploy our protocol on three GPU cloud servers and conduct experiments in both LAN and WAN environments. We choose CIFAR10/Tiny datasets and AlexNet/VGG16 models. With both 2-out-of-2 secret sharing~\footnote{Three parties are still involved, where two parties hold 2-out-of-2 secret shares and the other party acts as an assistant.} and replicated secret sharing, by using Bicoptor 2.0 DReLU protocol the E2E PPML inference performance is increased by a factor of 3-4 than that of P-Falcon~\cite{WWP22,piranha-code}, with less than 0.5\% accuracy loss. 
\end{itemize}  

\noindent \textbf{Paper Organization.} Section~\ref{sec:preliminary} introduces the settings, notations, and related backgrounds. By following the structure as illustrated in Fig.~\ref{fig:paperoverview}, the rest of the paper is organized as follows. In Section~\ref{sec:analysis} we conduct an in-depth theoretical analysis of the $e_1$ error in the probabilistic truncation protocol. Section~\ref{sec:e1acc} and~\ref{sec:drelu} are dedicated to the exploration of the implications of the $e_1$ error from both accuracy and efficiency perspectives. In Section~\ref{sec:e1acc}, we delve into the challenges posed by the $e_1$ error in the context of accuracy and present corresponding solutions. In Section~\ref{sec:drelu}, we discuss the impact of $e_1$ on efficiency and proposed deterministic truncation to mitigate its effects.  In Section~\ref{sec:evaluate}, we demonstrate the end-to-end PPML inference process using the optimized protocols and perform testing and evaluation. 

\section{Preliminaries} \label{sec:preliminary}
\subsection{System Settings} \label{sub:setting}
We mainly consider the unbalanced mode (UBL) of a three-party setting, i.e., two parties hold 2-out-of-2 secret shares and the other party acts as an assistant, which is similar to previous 3PC PPML works, e.g., Chameleon~\cite{RWT+-18}, ASTRA~\cite{CCP+-19}, QuantizedNN~\cite{AH19}, SecureNN~\cite{WGC19}, CryptFlow~\cite{RRK+-20}, BLAZE~\cite{PS20}, and SWIFT~\cite{KPP+-21}. This new DReLU protocol also works with replicated secret sharing (RSS). Our protocols can resist static (i.e., non-adaptive) and honest-but-curious (i.e., semi-honest) adversaries and are secure in the honest majority setting. This means that at most one of all three participants ($P_0$, $P_1$, and $P_2$) is honest-but-curious, i.e., no collusion between any two participants. We also assume that $P_0$ and $P_1$, $P_0$ and $P_2$, and $P_1$ and $P_2$ have pre-shared pseudorandom seeds $\textsf{seed}_{01}$, $\textsf{seed}_{02}$, and $\textsf{seed}_{12}$, respectively.

\subsection{Notations} \label{sub:notation}
We use $:=$ to denote the definition. Considering a secret input $x\in[0,2^\ell)$ is positive or negative, if $x\in[0,2^{\ell_x})$ or $x\in(2^\ell-2^{\ell_x},2^\ell)$, respectively. $\ell$ is the bit length of an element in $\mathbb{Z}_{2^\ell}$. $\ell_x$ is the precision bit length of $x$. $[x]$ refers to the shares of $x$. We use the following specific 2-out-of-2 secret sharing for all protocols in this paper:
$$[x]_0 :=x + R \ \text{mod}\ 2^\ell,\ [x]_1 :=-R\ \text{mod}\ 2^\ell,$$
and $x=[x]_0 + [x]_1\ \text{mod}\ 2^\ell$, $R$ is a random number belongs to $\mathbb{Z}_{2^\ell}$. $\xi$ is defined as the ``absolute value'' of $x$. Where $\xi:=x\ \text{mod}\ 2^\ell$ for positive $x$ and $\xi:=2^\ell-x\ \text{mod}\ 2^\ell$ for negative $x$.

The input $x$ is a fix-point number consisting of two parts: the fractional part and the integer part. The bit length of the fractional part is denoted by $\ell_x^\mathsf{frac}$, the bit length of the integer part is denoted by $\ell_x^\mathsf{int}$, and the total precision length of the input is denoted by $\ell_x := \ell_x^\mathsf{int} + \ell_x^\mathsf{frac}$. 

\begin{table}[ht]
  \caption{Notation table}
  \label{tab:notation}
  \centering
    \linespread{1.25}
    \footnotesize
      \begin{tabular*}{8.5cm}{@{\extracolsep{\fill}} p{1.5cm} p{7cm} }
        \toprule
        \textbf{Notation} & \textbf{Description}\\
        \hline
        $=$,$:=$ & equal to, define as\\
        \hline
        $P_0$,$P_1$,$P_2$ & $P_0$, $P_1$, and $P_2$ are three participants in our protocols.\\
        \hline
        $\textsf{seed}_{01}$, & $\textsf{seed}_{01}$, $\textsf{seed}_{02}$, $\textsf{seed}_{12}$ are pre-shared pseudorandom \\
        $\textsf{seed}_{02}$, & seeds among $P_0$ and $P_1$, $P_0$ and $P_2$, and $P_1$ and  \\
        $\textsf{seed}_{12}$  & $P_2$, respectively.\\
        \hline
        $x$,$\ell_x$,$\ell_x^\mathsf{int}$,$\ell_x^\mathsf{frac}$ 
        			   & An input $x$ with precision $\ell_x$. $x$ is defined as positive if \\
                       & $x\in [0,2^{\ell_x})$, and negative if $x\in (2^\ell-2^{\ell_x},2^\ell)\ \text{mod}\ 2^\ell$. \\
                       & $\ell_x:=\ell_x^\mathsf{int}+\ell_x^\mathsf{frac}$, where $\ell_x^\mathsf{int}$ and $\ell_x^\mathsf{frac}$ correspond to \\
                       & the binary integer precision and binary fraction \\
                       & precision, respectively.\\
        \hline $p$ & $p$ is a prime number, whose bit length is $\ell_x+1$.\\  
        \hline
        $[x]$ & The shares of $x$ in $\mathbb{Z}_{2^\ell}$.\\
        \hline
        $\xi$           & The ``absolute value'' of $x$. $\xi:=x$ if $x$ is positive.\\
                        & $\xi:=2^\ell - x\ \text{mod}\ 2^\ell$ if $x$ is negative.\\
        \hline
        $\mathsf{cut}$ & The cut operation does not preserve the sign.\\
        \hline
        $\mathsf{trc}$ & The truncation operation preserves the sign.\\
        \hline
        $\mathsf{cut}(\cdot,k)$ & $\mathsf{cut}(\cdot,k)$ cuts the last $k$ bits of the input.\\
        \hline
        $\mathsf{cut}(\cdot,k_1,k_2)$ & $\mathsf{cut}(\cdot,k_1,k_2)$ cuts the first $k_2$ bits and the last $k_1$ bits of\\
                      & the input.\\
        \hline
        $\mathsf{trc}(\cdot,k)$ 
                        & $\mathsf{trc}(\cdot,k)$ truncates the last $k$ bits of the input,\\
                        & and the results are elements in $\mathbb{Z}_{2^{\ell}}$.\\
        \hline
        $\overline{\mathsf{trc}}(\cdot,k)$ 
                        & New $\overline{\mathsf{trc}}(\cdot,k)$ truncates the last $k$ bits of the input,\\
                        & and the results are elements in $\mathbb{Z}_{2^{\ell-k}}$.\\
        \hline
        $\overline{\mathsf{trc}}(\cdot,k_1,k_2)$ & 
        New $\overline{\mathsf{trc}}(\cdot,k_1,k_2)$ truncates the first $k_2$ bits and the last \\
                        & $k_1$ bits of the input, and the results are elements\\
                        & in $\mathbb{Z}_{2^{\ell-k_1-k_2}}$.\\
        \hline
        $a,b,c,d,e$     & A preprocessed multiplication triple $[a],[b],[c]$,\\
                        & $d$ is the reconstructed value of $[x]-[a]\ \text{mod}\ 2^\ell$.\\
                        & $e$ is the reconstructed value of $[y]-[b]\ \text{mod}\ 2^\ell$.\\
                        & $[x\cdot y] = de + d[b] + e[a] + \gamma\ \text{mod}\ 2^\ell$.\\
        \hline
        $r,r^\prime:=\frac{r}{2^k}$    & A random pair used in the truncation protocol proposed \\
                        &in $\text{ABY}^3$. \\
        \hline
        $\alpha,\beta,\gamma$ & Constants.\\
        \hline 
        UBL & Three parties are involved, where two parties hold 2-out-of-2 \\
        & secret shares and the other party acts as an assistant.\\
        \hline 
        RSS & Three parties are involved, where each party holds \\
        & replicated secret shares.\\
        \bottomrule
	  \end{tabular*}
\end{table}

\subsection{Cut Function and Truncation Protocol} 
\noindent \textbf{The Cut Function.} The function $\mathsf{cut}(\alpha,k)$ is similar to the right shifting operation, which cuts the last $k$ bits of $\alpha$. Considering an integer $\alpha:=\sum_0^{\ell-1}\alpha_i\cdot2^i$, whose binary decomposition is $\alpha:=\{\alpha_{\ell-1},\cdots,\alpha_0\}$. Then, the binary form of the result is
$$\mathsf{cut}(\alpha,k):=\sum_k^{\ell-1}\alpha_i\cdot2^i=\{\alpha_{\ell-1},\cdots,\alpha_{k}\}.$$

We furthre define a function $\mathsf{cut}(\alpha,k_1,k_2)$, which cuts the first $k_2$ bits and the last $k_1$ of $\alpha$. The binary form of the result is
$$\mathsf{cut}(\alpha,k_1,k_2):=\sum_{k_1}^{\ell-k_2-1}\alpha_i\cdot2^i=\{\alpha_{\ell-k_2-1},\cdots,x_{k_1}\}.$$

\noindent \textbf{The Truncation Protocol.} Truncation is used to recover the fixed-point decimal precision after a multiplication operation, which is a key component in approximate computation. There are several truncation protocols proposed by researchers, which could be distinguished into two types, i.e., deterministic and probabilistic. The results of a truncation protocol are affected by two factors. One is a one-bit error, we name it $e_0$, and both deterministic and probabilistic truncation protocols carry this error. The other error $e_1$ can directly cause the failure in truncation. To avoid the error of $e_1$ happening frequently and leading to more serious problems, we usually choose a larger ring size. This also means that when using probabilistic truncation, we need to be careful in selecting parameters and there will be some limitations, e.g., ~\cite{MZ17,MR18}. 

We denote the truncation protocols proposed in SecureML~\cite{MZ17} and $\text{ABY}^3$~\cite{MR18} as $\mathsf{trc}(\cdot,k)$, which truncates the last $k$ bits of the input and its outputs are elements in $\mathbb{Z}_{2^{\ell}}$. This paper proposes two new truncation protocols: $\overline{\mathsf{trc}}(\cdot,k)$ which truncates the last $k$ bits of the input and its outputs are elements in $\mathbb{Z}_{2^{\ell-k}}$; $\overline{\mathsf{trc}}(\cdot,k_1,k_2)$ truncates the first $k_2$ bits and last $k_1$ of the input and the outputs are elements in $\mathbb{Z}_{2^{\ell-k_1-k_2}}$.

\subsection{The Truncation Protocols Proposed in Prior Work Used in This Paper} 
The non-interactive probabilistic truncation protocol proposed in SecureML~\cite{MZ17} is summarised in Alg.~\ref{alg:smltrc}, and its correctness illustrated by Theorem~\ref{thm:smltrc} is proved in~\cite[Sect.~4.1]{MZ17}.

\begin{algorithm}[ht]
  \textbf{Input}: shares of $x\in[0,2^{\ell_x})\bigcup(2^\ell-2^{\ell_x},2^\ell)$ in $\mathbb{Z}_{2^\ell}$, number of bits to be truncated $k$\\
  \textbf{Output}: shares of $\mathsf{trc}(x,k)$ in $\mathbb{Z}_{2^\ell}$
      \begin{algorithmic}[1]
        \State $P_0$ sets $[\mathsf{trc}(x,k)]_0:=\mathsf{cut}([x]_0,k)\ \text{mod}\ 2^\ell$.
        \State $P_1$ sets $[\mathsf{trc}(x,k)]_1:=2^\ell - \mathsf{cut}(2^\ell - [x]_1,k)\ \text{mod}\ 2^\ell$.
      \end{algorithmic}
    \caption{The Truncation Protocol Proposed in SecureML~\cite{MZ17}.}
    \label{alg:smltrc}
\end{algorithm}

Theorem~\ref{thm:smltrc} describes the occurrence of $e_0$. We define the ``slack''~\cite{MRV+-21} as $\ell - \ell_x$. While invoking Alg.~\ref{alg:smltrc} in an MPC-based PPML protocol, we should choose a larger ring size to ensure enough slack.

\begin{theorem} \label{thm:smltrc}
  In a ring $\mathbb{Z}_{2^\ell}$, let $x\in[0,2^{\ell_x})\bigcup(2^\ell-2^{\ell_x},2^\ell)$, where $\ell>\ell_x + 1$. Then the outputs of Alg.~\ref{alg:smltrc} satisfy the following results with probability $1-\frac{1}{2^{\ell-\ell_x-1}}$, where $\mathsf{bit}:=\{0,1\}$. 
  \begin{itemize}[leftmargin=*]
      \item For a positive $x$, $\mathsf{trc}(x,k)=\mathsf{cut}(\xi,k) + \mathsf{bit}$.
      \item For a negative $x$, $\mathsf{trc}(x,k)=2^\ell-\mathsf{cut}(\xi,k) - \mathsf{bit}$
  \end{itemize}
\end{theorem}

$\text{ABY}^3$~\cite{MR18} (Alg.~\ref{alg:aby3trc}) proposes an $n$-party interactive probabilistic truncation protocol. The participants first reconstruct the masked input, and locally truncate the opened value. Then, the participants remove the mask using a pre-shared element and obtain the final result. The protocol is summarised in Alg.~\ref{alg:aby3trc}, note that Alg.~\ref{alg:aby3trc} is also probabilistic and hence constrained by the slack. 

\begin{algorithm}[ht]
  \textbf{Preprocessing}: the shares of $r^\prime:=\frac{r}{2^k}$\\
  \textbf{Input}: shares of $x\in[0,2^{\ell_x})\bigcup(2^\ell-2^{\ell_x},2^\ell)$ in $\mathbb{Z}_{2^\ell}$, number of bits to be truncated $k$\\
  \textbf{Output}: shares of $\mathsf{trc}(x,k)$ in $\mathbb{Z}_{2^\ell}$
      \begin{algorithmic}[1]
        \State $P_i$ reconstructs $\alpha$ from $[x]_i + [r]_i$.\footnotemark
        \State $P_i$ sets $[\mathsf{trc}(x,k)]_i:=\frac{\alpha}{2^k} - [r^\prime]$.
      \end{algorithmic}
    \caption{The Truncation Protocol Proposed in $\text{ABY}^3$~\cite{MR18}.}
    \label{alg:aby3trc}
\end{algorithm}
\footnotetext{The truncation protocol proposed in $\text{ABY}^3$~\cite{MR18} uses $[x]_i - [r]_i$ for $\alpha$ and then adds $[r^\prime]$ to retrive $[\mathsf{trc}(x,k)]_i$, but we believe this is a typo.}

As mentioned in previous sections, the truncation protocols of both SecureML~\cite{MZ17} and $\text{ABY}^3$~\cite{MR18} are probabilistic, and therefore, they both suffer from $e_1$. We have already discussed the serious consequences of $e_1$. In the next sections, we will analyze in detail how $e_1$ is generated and whether we can fundamentally eliminate it with other methods.

\section{In-depth Analysis of $e_1$ in Probabilistic Truncation Protocols} \label{sec:analysis}
In this section, we will conduct an in-depth analysis of $e_1$ in the truncation protocols proposed in SecureML~\cite{MZ17} and $\text{ABY}^3$~\cite{MR18} (Alg.~\ref{alg:smltrc} and Alg.~\ref{alg:aby3trc}). We first introduce a more fundamental cut function to better understand the nature of $e_1$, and then explain how this fundamental cut function envolves in these truncation protocols.

\noindent \textbf{A Fundamental Cut Function} 
In our analysis, we find that truncation protocols are based on combinations or variations of cut functions. Since the essence of the occurrence of $e_1$ is also on the cut function, we first formally define the cut function and provide the key properties of the cut function in Theorem~\ref{thm:cut} to better understand the generation of $e_1$ and to better understand the essence of truncation protocols.

\begin{definition}
  For $\alpha,\beta \in \mathbb{Z}_{2^\ell}$, we define $\mathsf{LT}(\alpha,\beta):=1$ iff $\alpha<\beta$, and $0$ otherwise.
\end{definition}

\begin{theorem} \label{thm:cut} 
  For $\alpha,\beta \in \mathbb{Z}_{2^\ell}$ and $\mathsf{bit} := \{0,1\}$,
  \begin{itemize} [leftmargin=*]
    \item $\mathsf{cut}(\alpha+\beta\ \text{mod}\ 2^\ell,k) = \mathsf{cut}(\alpha,k) + \mathsf{cut}(\beta,k) - \mathsf{LT}(\alpha+\beta\ \text{mod}\ 2^\ell,a)\cdot\mathsf{cut}(2^\ell,k) + \mathsf{bit}$
    \item $\mathsf{cut}(\alpha-\beta\ \text{mod}\ 2^\ell,k) = \mathsf{cut}(\alpha,k) - \mathsf{cut}(\beta,k) + \mathsf{LT}(\alpha, \alpha-\beta\ \text{mod}\ 2^\ell)\cdot\mathsf{cut}(2^\ell,k) - \mathsf{bit}$
  \end{itemize}
\end{theorem}

The proof of Theorem~\ref{thm:cut} could be found in Appendix~\ref{app:sub:proof:thm:cut}. 

We recall that the truncation protocol of SecureML~\cite{MZ17} transfers the truncation operation on plaintext $x$ into the cut operations on shares $[x]_0 := x + R\ \text{mod}\ 2^\ell$ and $[x]_1 := -R\ \text{mod}\ 2^\ell$, i.e., $$\mathsf{trc}(x,k) = \mathsf{cut}([x]_0,k)-\mathsf{cut}(2^\ell - [x]_1,k)\ \text{mod}\ 2^\ell.$$ By substituting $[x]_0 = x + R\ \text{mod}\ 2^\ell$ and $[x]_1= -R\ \text{mod}\ 2^\ell$, Theorem~\ref{thm:smltrc} and Theorem~\ref{thm:cut}, we obtain Corollary~\ref{clr:cut2}.
\begin{corollary} \label{clr:cut2}
	In a ring $\mathbb{Z}_{2^\ell}$, let $x\in[0,2^{\ell_x})\bigcup(2^\ell-2^{\ell_x},2^\ell)$, where $\ell>\ell_x + 1$. Then the outputs of Alg.~\ref{alg:smltrc} satisfy the following results, where $\mathsf{bit}:=\{0,1\}$. 
	\begin{itemize} [leftmargin=*]
	  \item For a positive $x$, $\mathsf{trc}(x,k)=\mathsf{cut}(x,k)-\mathsf{LT}(x+R\ \text{mod}\ 2^\ell,x)\cdot\mathsf{cut}(2^\ell,k)+\mathsf{bit}\ \text{mod}\ 2^\ell$.
	  \item For a negative $x$, $\mathsf{trc}(x,k)=-\mathsf{cut}(-x,k)+\mathsf{LT}(x,x+R\ \text{mod}\ 2^\ell)\cdot\mathsf{cut}(2^\ell,k)-\mathsf{bit}\ \text{mod}\ 2^\ell$.
	\end{itemize}
\end{corollary}

For positive $x$, we expect $\mathsf{trc}(x,k) = \mathsf{cut}(x,k)+\mathsf{bit}\ \text{mod}\ 2^\ell$. However, we observe that there is an additional term $\mathsf{LT}(x+R\ \text{mod}\ 2^\ell,x)\cdot\mathsf{cut}(2^\ell,k)$ in Corollary~\ref{clr:cut2}. When $x + R\ \text{mod}\ 2^\ell > x$, $\mathsf{LT}(x+R\ \text{mod}\ 2^\ell,x) = 0$, this extra term disappears, which makes $\mathsf{trc}(x,k) = \mathsf{cut}(x,k)$ consistent with our expectation. When $x + R\ \text{mod}\ 2^\ell < x$, $\mathsf{LT}(x+R\ \text{mod}\ 2^\ell,x) = 1$, the error term exists causing the failure of truncation, also known as $e_1$. Similarly, we expect $\mathsf{trc}(x,k)=-\mathsf{cut}(-x,k)-\mathsf{bit}\ \text{mod}\ 2^\ell$ for negative $x$. $e_1$ occurs when $\mathsf{LT}(x,x+R\ \text{mod}\ 2^\ell)=1$. Hence, we claim that 
\begin{align*}
	&\mathsf{P}(\text{SecureML truncation failure}) = \mathsf{P}(e_1)\\
	=&\mathsf{P}(x + R\ \text{mod}\ 2^\ell < x\ |\ x\in[0,2^{\ell_x}))\\
	=&\mathsf{P}(x < x + R\ \text{mod}\ 2^\ell\ |\ x\in(2^\ell-2^{\ell_x},2^\ell)) \\
  =& \frac{1}{2^{\ell-\ell_x-1}}.
\end{align*}
We further notice that, the larger the difference between $\ell$ and $\ell_x$, the lower the probability of failure in truncations. 

The truncation protocol of $\text{ABY}^3$~\cite{MR18} (Alg.~\ref{alg:aby3trc}) also suffers from $e_1$ as the truncation protocol of SecureML~\cite{MZ17} (Alg.~\ref{alg:smltrc}), and the underlying cause of the errors and its probability are the same i.e., 
\begin{align*}
	&\mathsf{P}(\text{ABY}^3\ \text{truncation failure}) = \mathsf{P}(e_1)\\
	=&\mathsf{P}(\alpha = x + r\ \text{mod}\ 2^\ell < x\ |\ x\in[0,2^{\ell_x}))\\
	=&\mathsf{P}(x < \alpha = x + r\ \text{mod}\ 2^\ell\ |\ x\in(2^\ell-2^{\ell_x},2^\ell)) \\
  =& \frac{1}{2^{\ell-\ell_x-1}}.
\end{align*}

In conclusion, minimizing the probability of the occurrence of $e_1$ is equivalent to maximizing $\ell - \ell_x$. Hence, the parameter $\ell_x$ should be sufficiently smaller than $\ell$, i.e., $\ell_x \ll \ell$. We would like to emphasize that the above conclusion is based on an assumption that $R$, $r$ are uniformly and randomly chosen from the ring, which is also a necessary condition for security. 

\begin{table}[t!]
  \caption{Accuracy when applying randoms in $\text{ABY}^3$ for truncation and applying the truncate-then-multiply solution in Piranha PPML inference implementations~\cite{WWP22,piranha-code}, including P-SecureML (2-Party), P-Falcon (3-Party) and P-Fantastic (4-Party). Entries with (f) indicate the use of fixed numbers, while entries with (r) indicate the use of random numbers.}
  \label{tab:accuracy}
  \centering
  \linespread{1.25}
  \footnotesize
\begin{tabular*}{8.5cm}{@{\extracolsep{\fill}} p{1cm} l p{1.55cm} p{1.1cm} p{1.4cm} } 
  \toprule
  \multirow{2}*{\textbf{Model}} & \multirow{2}*{\textbf{Type}} 
  &  \multicolumn{3}{c}{\textbf{Fraction precision $\ell_x^\mathsf{frac}=26$}} \\ 
  &
  &\textbf{P-SecureML} & \textbf{P-Falcon} & \textbf{P-Fantastic} \\ 
  \hline
  \multirow{3}{0.9cm}{\centering CIFAR10\_\\AlexNet}
    & mult-then-trc (f)  & 69.63\% & 69.63\% & 69.63\% \\ 
    & mult-then-trc (r) & 12.74\% & 12.73\% & 12.74\% \\ 
    & trc-then-mult (r) & 69.62\% & 69.64\% & 69.62\% \\ 
  \hline
  \multirow{3}{0.9cm}{\centering Tiny\_\\AlexNet} 
    & mult-then-trc (f)  & 26.39\% & 26.39\% & 26.39\% \\ 
    & mult-then-trc (r) & 0.45\% & 0.44\% & 0.45\% \\ 
    & trc-then-mult (r) & 26.35\% & 26.35\% & 26.35\% \\ 
  \hline
  \multirow{3}{0.9cm}{\centering CIFAR10\_\\VGG16}
    & mult-then-trc (f)  & 88.31\% & 88.31\% & 88.31\% \\ 
    & mult-then-trc (r) & 10.60\% & 9.89\% & 10.60\% \\ 
    & trc-then-mult (r) & 88.29\% & 88.29\% & 88.35\% \\ 
  \hline
  \multirow{3}{0.9cm}{\centering Tiny\_\\VGG16} 
    & mult-then-trc (f) & 54.89\% & 54.89\% & 54.89\% \\ 
    & mult-then-trc (r) & 0.43\% & 0.41\% & 0.50\% \\ 
    & trc-then-mult (r) & 54.92\% & 54.92\% & 54.88\% \\ 
  \bottomrule
  \end{tabular*}
\end{table}

\section{Impact of $e_1$ on Accuracy in PPML and Propoesed Solutions} \label{sec:e1acc}

Based on the previous analysis, we understand that the occurrence of $e_1$ can cause a substantial deviation from the desired result, which can further affect the training and inference accuracy. We would like to point out that some existing works have used fixed numbers instead of random numbers to simplify their implementation, which inadvertently hides $e_1$. 

In Sect.~\ref{sub:bug}, we will provide a comprehensive explanation of why the use of fixed numbers instead of random numbers conceals the manifestation of $e_1$. Furthermore, we will present the actual inference accuracy when employing random numbers under the parameter recommendations provided in these works.

In Sect.~\ref{sub:linear}, we propose a solution that enables the selection of larger $\ell_x$ values while employing random numbers to enhance inference accuracy.

\subsection{The Implementation Bug Related to $e_1$ in Existing Works} \label{sub:bug}
Some existing works fail to discover that the parameters they select are insufficient to support the correctness of inference, as they use fixed numbers instead of random numbers in their truncation protocols. 

Looking purely from the perspective that $\ell_x$ should be much smaller than $\ell$, existing work has indeed chosen appropriate parameters. For example, in Piranha~\cite{WWP22,piranha-code}, all P-SecureML, P-Falcon, and P-Fantastic implementations invoke the truncation protocol of $\text{ABY}^3$~\cite{MR18} (Alg.~\ref{alg:aby3trc}) and the recommended precision $\ell_x^\mathsf{frac}=26$ and $\ell=64$ which does satisfy $\ell_x \ll \ell$. However, after a single multiplication operation, the size of $\ell_x$ is doubled. The new $\ell^\prime_x = 2\cdot\ell_x = 62$ if $\ell_x^\mathsf{int}=5$, in which $\ell^\prime_x$ is very close to $\ell=64$. Theoretically, this would result in a very high probability of $e_1$ subsequently would lead to a decrease in inference accuracy. However, based on the experimental results of the aforementioned work, this does not occur. 

The reason for this is that the above works fix $r$ to be $2^{26}$ in Alg.~\ref{alg:aby3trc}. In this case, for positive $x$, $$\mathsf{P}(e_1)=\mathsf{P}(\alpha = x + r\ \text{mod}\ 2^\ell < x | x\in[0,2^{\ell_x}))=0.~\footnote{If $x$ is a positive 62-bit long number and $r$ is $2^{26}$, $x+r\ \text{mod}\ 2^{26}$ will never be small than $x$.}$$ Therefore, $e_1$ does not occur as expected. We replace the fixed $r$ in Alg.~\ref{alg:aby3trc} with random numbers and rerun the experiments, and the results are exhibited in Fig.~\ref{fig:aby3trc} and Tab.~\ref{tab:accuracy}.

Fig.~\ref{fig:aby3trc} illustrates the impact of using random numbers(r) and fixed numbers(f) on inference accuracy under various models and parameters. When $\ell_x$ is getting close to $\ell$, $\mathsf{P}(e_1)$ increases and further leads to decreasing in inference accuracy. Fig.~\ref{fig:aby3trc}(a)-(d) illustrates the scenarios where $e_1$ may occur in a practical situation, that is, when random numbers are employed. It demonstrates that selecting the parameter combination $\ell_x^\mathsf{frac} = 26$ and $\ell=64$, as claimed Piranha~\cite{WWP22,piranha-code}, is not suitable. Similarly, Fig.~\ref{fig:aby3trc}(e) shows that $\ell_x^\mathsf{frac} = 13$ and $\ell=32$ is also inappropriate. In addition, we demonstrate the impact of using fixed or random numbers on the inference accuracy for the 2-party, 3-party, and 4-party protocols when selecting $\ell_x^\mathsf{frac} = 26$ in Tab.~\ref{tab:accuracy}.

\subsection{Proposed Solution to Address the Implementation Bug Related to $e_1$} \label{sub:linear}

\begin{figure}[t] 
  \centering 
      \begin{tikzpicture}
      \footnotesize
      \begin{axis}[
        height=3.5cm, width= 8.5cm, axis lines=middle,
        axis line style={->}, xmin=0, xmax=30,
        ymin=0, ymax=1,
        x label style={at={(axis description cs:0.5,-0.2)},anchor=north},
        y label style={at={(axis description cs:-0.1,0.5)},rotate=90,anchor=south},
        title={(a) P-Falcon experiments in $\mathbb{Z}_{2^{64}}$ for CIFAR10\_AlexNet},
        title style={at={(axis description cs:0.5,0.9)},anchor=north},
          xlabel=Fraction precision (bit),
          ylabel=Inference accuracy,
          legend style={at={(1.05,1.4)},anchor=east},
          legend columns=2]
      \addplot[smooth,mark=-,gray] plot coordinates {
          (1,0.10) (2,0.10) (3,0.10) (4,0.10) (5,0.10)
          (6,0.10) (7,0.10) (8,0.10) (9,0.10) (10,0.10)
          (11,0.10) (12,0.10) (13,0.10) (14,0.10) (15,0.10)
          (16,0.10) (17,0.10) (18,0.10) (19,0.10) (20,0.10)
          (21,0.10) (22,0.10) (23,0.10) (24,0.10) (25,0.10)
          (26,0.10)
      };
      \addlegendentry{Random Guessing}
      \addplot[smooth,mark=*,blue] plot coordinates {
          (1,0.0938) (2,0.0989) (3,0.0982) (4,0.0984) (5,0.244400)
          (6,0.600000) (7,0.677400) (8,0.692800) (9,0.695000) (10,0.696100)
          (11,0.695800) (12,0.6963) (13,0.6963) (14,0.6960) (15,0.6965)
          (16,0.6964) (17,0.6962) (18,0.6962) (19,0.6962) (20,0.696200)
          (21,0.695200) (22,0.690400) (23,0.668700) (24,0.586200) (25,0.376200)
          (26,0.127300)
      };
      \addlegendentry{$\text{ABY}^3$ mult-then-trc (r)}
      \addplot[smooth,color=red,mark=x]
          plot coordinates {
            (1,0.10) (2,0.10) (3,0.10) (4,0.10) (5,0.1698)
            (6,0.5602) (7,0.6704) (8,0.6902) (9,0.6948) (10,0.6965)
            (11,0.6958) (12,0.6964) (13,0.6958) (14,0.6964) (15,0.6963)
            (16,0.6963) (17,0.6963) (18,0.6963) (19,0.6963) (20,0.6963)
            (21,0.6963) (22,0.6963) (23,0.6963) (24,0.6963) (25,0.6963)
            (26,0.6963)
          };
      \addlegendentry{$\text{ABY}^3$ mult-then-trc (f)}
      \addplot[smooth,color=green,mark=triangle*]
          plot coordinates {
            (1,0.10) (2,0.10) (3,0.10) (4,0.10) (5,0.191900)
            (6,0.499700) (7,0.532200) (8,0.660900) (9,0.657800) (10,0.684900)
            (11,0.684900) (12,0.690300) (13,0.691700) (14,0.695000) (15,0.693600)
            (16,0.696000) (17,0.695000) (18,0.695500) (19,0.695600) (20,0.696000)
            (21,0.696500) (22,0.696300) (23,0.696500) (24,0.696200) (25,0.696500)
            (26,0.696400)
          };
      \addlegendentry{$\text{ABY}^3$ trc-then-mult (r)}
      \end{axis}
      \end{tikzpicture}
      \begin{tikzpicture}
        \footnotesize
        \begin{axis}[
          height=3.5cm, width= 8.5cm, axis lines=middle,
          axis line style={->}, xmin=0, xmax=30,
          ymin=0, ymax=1.2,
          x label style={at={(axis description cs:0.5,-0.2)},anchor=north},
          y label style={at={(axis description cs:-0.1,0.5)},rotate=90,anchor=south},
          title={(c) P-Falcon experiments in $\mathbb{Z}_{2^{64}}$ for CIFAR10\_VGG16},
          title style={at={(axis description cs:0.5,0.9)},anchor=north},
            xlabel=Fraction precision (bit),
            ylabel=Inference accuracy]
        \addplot[smooth,mark=-,gray] plot coordinates {
            (1,0.10) (2,0.10) (3,0.10) (4,0.10) (5,0.10)
            (6,0.10) (7,0.10) (8,0.10) (9,0.10) (10,0.10)
            (11,0.10) (12,0.10) (13,0.10) (14,0.10) (15,0.10)
            (16,0.10) (17,0.10) (18,0.10) (19,0.10) (20,0.10)
            (21,0.10) (22,0.10) (23,0.10) (24,0.10) (25,0.10)
            (26,0.10)
        };
        \addplot[smooth,mark=*,blue] plot coordinates {
            (1,0.0938) (2,0.0989) (3,0.0982) (4,0.099800) (5,0.096800)
            (6,0.242500) (7,0.805800) (8,0.872800) (9,0.879700) (10,0.8831)
            (11,0.8831) (12,0.883100) (13,0.883100) (14,0.883200) (15,0.882900)
            (16,0.8831) (17,0.8831) (18,0.8831) (19,0.8831) (20,0.868200)
            (21,0.825500) (22,0.682800) (23,0.347800) (24,0.108300) (25,0.099900)
            (26,0.098900)
        };
        \addplot[smooth,color=red,mark=x]
            plot coordinates {
              (1,0.10) (2,0.10) (3,0.10) (4,0.10) (5,0.100000)
              (6,0.148500) (7,0.760400) (8,0.869500) (9,0.879600) (10,0.881400)
              (11,0.881800) (12,0.883100) (13,0.883300) (14,0.883100) (15,0.883100)
              (16,0.882900) (17,0.883000) (18,0.883100) (19,0.883100) (20,0.883100)
              (21,0.883100) (22,0.883100) (23,0.883100) (24,0.883100) (25,0.883100)
              (26,0.883100)
            };
        \addplot[smooth,color=green,mark=triangle*] plot coordinates {
          (1,0.10) (2,0.10) (3,0.10) (4,0.10) (5,0.100000)
          (6,0.120300) (7,0.168200) (8,0.711600) (9,0.730400) (10,0.857300)
          (11,0.858200) (12,0.879200) (13,0.8831) (14,0.8831) (15,0.8831)
          (16,0.8831) (17,0.8831) (18,0.8831) (19,0.8831) (20,0.8831)
          (21,0.8831) (22,0.8831) (23,0.8831) (24,0.883100) (25,0.8831)
          (26,0.883100)
        };
        \end{axis}
      \end{tikzpicture}
      \begin{tikzpicture}
        \footnotesize
        \begin{axis}[
          height=3.5cm, width= 8.5cm, axis lines=middle,
          axis line style={->}, xmin=0, xmax=30,
          ymin=0, ymax=0.4,
          x label style={at={(axis description cs:0.5,-0.2)},anchor=north},
          y label style={at={(axis description cs:-0.1,0.5)},rotate=90,anchor=south},
          title={(b) P-Falcon experiments in $\mathbb{Z}_{2^{64}}$ for Tiny\_AlexNet},
          title style={at={(axis description cs:0.5,0.9)},anchor=north},
            xlabel=Fraction precision (bit),
            ylabel=Inference accuracy]
        \addplot[smooth,mark=-,gray] plot coordinates {
            (1,0.005) (2,0.005) (3,0.005) (4,0.005) (5,0.005)
            (6,0.005) (7,0.005) (8,0.005) (9,0.005) (10,0.005)
            (11,0.005) (12,0.005) (13,0.005) (14,0.005) (15,0.005)
            (16,0.005) (17,0.005) (18,0.005) (19,0.005) (20,0.005)
            (21,0.005) (22,0.005) (23,0.005) (24,0.005) (25,0.005)
            (26,0.005)
        };
        \addplot[smooth,mark=*,blue] plot coordinates {
            (1,0.004600) (2,0.004600) (3,0.004600) (4,0.005000) (5,0.017800)
            (6,0.146900) (7,0.243800) (8,0.259300) (9,0.263400) (10,0.263800)
            (11,0.263200) (12,0.264000) (13,0.263400) (14,0.263800) (15,0.263600)
            (16,0.263900) (17,0.263800) (18,0.263700) (19,0.262800) (20,0.260500)
            (21,0.250200) (22,0.213100) (23,0.112700) (24,0.018300) (25,0.005400)
            (26,0.004400)
        };
        \addplot[smooth,color=red,mark=x] plot coordinates {
            (1,0.005) (2,0.005) (3,0.005) (4,0.005) (5,0.014300)
            (6,0.138100) (7,0.242100) (8,0.259500) (9,0.263500) (10,0.262900)
            (11,0.263300) (12,0.264100) (13,0.263600) (14,0.264200) (15,0.263900)
            (16,0.263900) (17,0.263900) (18,0.263800) (19,0.263800) (20,0.263800)
            (21,0.263800) (22,0.263800) (23,0.263800) (24,0.263800) (25,0.263800)
            (26,0.263800)
        };
        \addplot[smooth,color=green,mark=triangle*] plot coordinates {
            (1,0.005) (2,0.005) (3,0.005) (4,0.005) (5,0.009000)
            (6,0.050300) (7,0.052200) (8,0.193000) (9,0.197800) (10,0.248500)
            (11,0.247000) (12,0.261300) (13,0.2639) (14,0.2639) (15,0.2639)
            (16,0.2639) (17,0.2639) (18,0.2639) (19,0.2639) (20,0.2639)
            (21,0.2639) (22,0.2639) (23,0.2639) (24,0.2639) (25,0.2639)
            (26,0.263800)
          };
        \end{axis}
      \end{tikzpicture}
      \begin{tikzpicture}
        \footnotesize
        \begin{axis}[
          height=3.5cm, width= 8.5cm, axis lines=middle,
          axis line style={->}, xmin=0, xmax=30,
          ymin=0, ymax=0.8,
          x label style={at={(axis description cs:0.5,-0.2)},anchor=north},
          y label style={at={(axis description cs:-0.1,0.5)},rotate=90,anchor=south},
          title={(d) P-Falcon experiments in $\mathbb{Z}_{2^{64}}$ for Tiny\_VGG16},
          title style={at={(axis description cs:0.5,0.9)},anchor=north},
            xlabel=Fraction precision (bit),
            ylabel=Inference accuracy]
        \addplot[smooth,mark=-,gray] plot coordinates {
            (1,0.005) (2,0.005) (3,0.005) (4,0.005) (5,0.005)
            (6,0.005) (7,0.005) (8,0.005) (9,0.005) (10,0.005)
            (11,0.005) (12,0.005) (13,0.005) (14,0.005) (15,0.005)
            (16,0.005) (17,0.005) (18,0.005) (19,0.005) (20,0.005)
            (21,0.005) (22,0.005) (23,0.005) (24,0.005) (25,0.005)
            (26,0.005)
        };
        \addplot[smooth,mark=*,blue] plot coordinates {
            (1,0.004600) (2,0.004600) (3,0.004600) (4,0.005600) (5,0.015800)
            (6,0.412600) (7,0.528100) (8,0.543900) (9,0.549000) (10,0.549900)
            (11,0.549700) (12,0.549400) (13,0.548600) (14,0.548300) (15,0.5489)
            (16,0.5489) (17,0.5489) (18,0.5489) (19,0.513200) (20,0.422300)
            (21,0.191900) (22,0.012000) (23,0.004800) (24,0.004000) (25,0.005600)
            (26,0.004100)
        };
        \addplot[smooth,color=red,mark=x] plot coordinates {
            (1,0.005) (2,0.005) (3,0.005) (4,0.005) (5,0.0161)
            (6,0.4144) (7,0.5282) (8,0.5434) (9,0.5492) (10,0.5495)
            (11,0.5496) (12,0.5490) (13,0.5487) (14,0.5487) (15,0.5487)
            (16,0.5487) (17,0.5487) (18,0.5487) (19,0.5487) (20,0.5487)
            (21,0.5487) (22,0.5487) (23,0.5487) (24,0.5487) (25,0.5487)
            (26,0.5487)
        };
        \addplot[smooth,color=green,mark=triangle*] plot coordinates {
            (1,0.005) (2,0.005) (3,0.005) (4,0.005) (5,0.009000)
            (6,0.005900) (7,0.005200) (8,0.132800) (9,0.140000) (10,0.453500)
            (11,0.457200) (12,0.532100) (13,0.5489) (14,0.5489) (15,0.5489)
            (16,0.5489) (17,0.5489) (18,0.5489) (19,0.5489) (20,0.5489)
            (21,0.5489) (22,0.5489) (23,0.5489) (24,0.5489) (25,0.5489)
            (26,0.5489)
        };
        \end{axis}
      \end{tikzpicture}
      \begin{tikzpicture}
        \footnotesize
        \begin{axis}[
          height=3.5cm, width= 8.5cm, axis lines=middle,
          axis line style={->}, xmin=0, xmax=15,
          ymin=0, ymax=1,
          x label style={at={(axis description cs:0.5,-0.2)},anchor=north},
          y label style={at={(axis description cs:-0.1,0.5)},rotate=90,anchor=south},
          title={(e) P-Falcon experiments in $\mathbb{Z}_{2^{32}}$ for Tiny\_VGG16},
          title style={at={(axis description cs:0.5,0.9)},anchor=north},
            xlabel=Fraction precision (bit),
            ylabel=Inference accuracy,
            legend style={at={(1.05,0.7)},anchor=east}, legend columns=3]
        
        \addplot[smooth,mark=-,gray] plot coordinates {
          (1,0.005) (2,0.005) (3,0.005) (4,0.005) (5,0.005)
          (6,0.005) (7,0.005) (8,0.005) (9,0.005) (10,0.005)
          (11,0.005) (12,0.005) (13,0.005) 
        };
        \addplot[smooth,mark=*,blue] plot coordinates {
            (1,0.005000) (2,0.005000) (3,0.005000) (4,0.004900) (5,0.012600)
            (6,0.053400) (7,0.005300) (8,0.005) (9,0.005) (10,0.005000)
            (11,0.00500) (12,0.00500) (13,0.00500) 
        };
        \addplot[smooth,color=red,mark=x]
            plot coordinates {
              (1,0.005000) (2,0.005000) (3,0.005000) (4,0.005000) (5,0.016100)
              (6,0.414400) (7,0.528200) (8,0.543400) (9,0.549200) (10,0.549500)
              (11,0.549600) (12,0.547300) (13,0.030800) 
            };
        \end{axis}
      \end{tikzpicture}
  \caption{Comparison of the effect of different precisions with fixed/random numbers on inference accuracy for CIFAR10\_AlexNet, Tiny\_AlexNet, CIFAR10\_VGG16, and Tiny\_VGG16. The experiments are carried using ring $\mathbb{Z}_{2^{64}}$ or $\mathbb{Z}_{2^{32}}$ in $\text{ABY}^3$ truncation protocol in P-Falcon~\cite{WWP22,piranha-code}. Entries with (f) indicate the use of fixed numbers, while entries with (r) indicate the use of random numbers.}
  \label{fig:aby3trc}
  \end{figure}
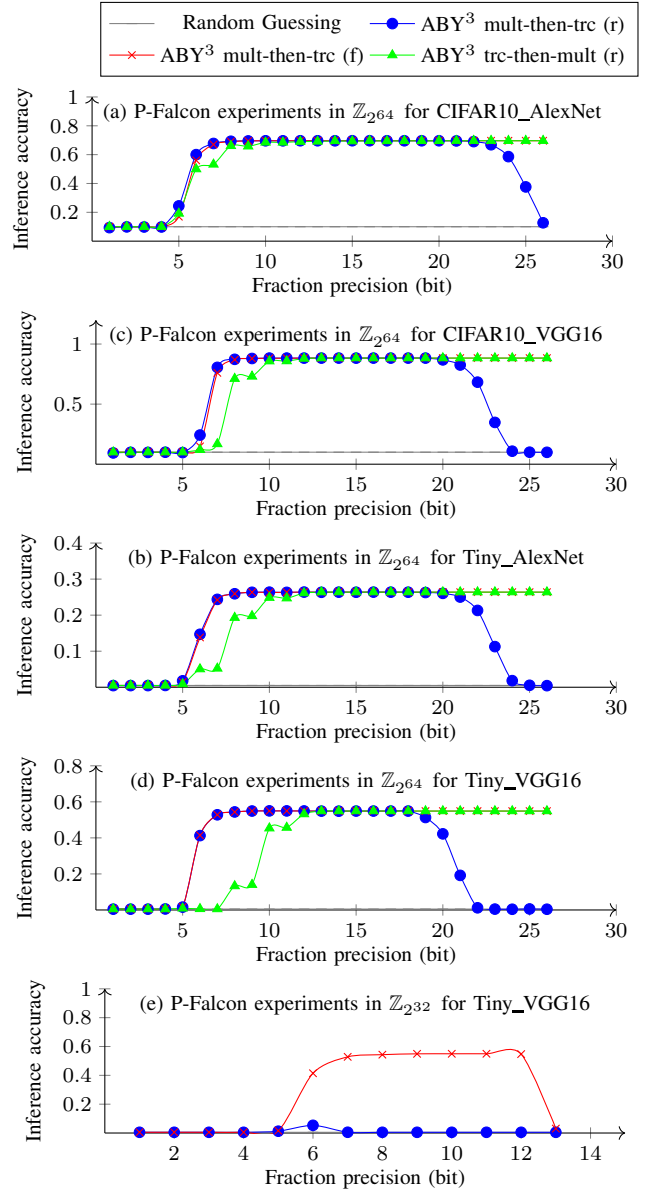

To address the issue of $e_1$ occurring due to the product of two numbers approaching the length of $\ell$ and thereby affecting inference accuracy, an intuitive solution is to reduce the precision of both numbers before performing multiplication. In other words, we propose ``truncate-then-multiply'' as a replacement for ``multiply-then-truncate''. Our solution is described in Alg.~\ref{alg:linear}. Through experiments, we demonstrate that under the original parameter selection, using the  ``truncate-then-multiply'' approach achieves inference accuracy equivalent to plaintext inference. 
\begin{algorithm}[ht]
  \textbf{Input}: shares of $x,y\in[0,2^{\ell_x})\bigcup(2^\ell-2^{\ell_x},2^\ell)$ in $\mathbb{Z}_{2^\ell}$\\
  \textbf{Output}: shares of $\mathsf{trc}(x\cdot y,\ell_x^\mathsf{frac})$ in $\mathbb{Z}_{2^\ell}$
      \begin{algorithmic}[1]
        \State $P_i$ runs Alg.~\ref{alg:smltrc} or Alg.~\ref{alg:aby3trc} to get $[\mathsf{trc}(x,\frac{\ell_x^\mathsf{frac}}{2})]$
        \Statex and $[\mathsf{trc}(y,\frac{\ell_x^\mathsf{frac}}{2})]$.
        \State $P_i$ computes $[\mathsf{trc}(x\cdot y,\ell_x^\mathsf{frac})]=[\mathsf{trc}(x,\frac{\ell_x^\mathsf{frac}}{2})\cdot\mathsf{trc}(y,\frac{\ell_x^\mathsf{frac}}{2})]$.
      \end{algorithmic}
    \caption{Truncate-then-multiply Solution.}
    \label{alg:linear}
\end{algorithm}

The triangle (green) line in Fig.~\ref{fig:aby3trc} demonstrates that our solution prevents the inference accuracy from dropping sharply due to high precision when using random numbers in all models. Tab.~\ref{tab:accuracy} also shows the same results, indicating that our solution effectively addresses the original issue of linear layer parameter limitation.

We observe some different patterns in the triangle (green) line compared to the other two lines, one being the staircase-like pattern when the triangle (green) line is increasing, and the other being that in Fig.~\ref{fig:aby3trc}(d), the inference accuracy only converged at precision 12, whereas the other two lines converged at around 7. We further analyze these two phenomena and find that the staircase pattern occurred because when the precision is odd, we cannot truncate the same number of bits for both numbers to be multiplied. For example, when the precision is $7$, we need to truncate $3$ bits for one number and $4$ bits for the other. As for the slower-than-expected increase in inference accuracy in Fig.~\ref{fig:aby3trc}(d), it was due to the fact that in a matrix multiplication in the CNN layer, there are multiple multiplications, and we do not perform resharing after each multiplication. Instead, we accumulate and then reshare after all multiplications in a matrix multiplication are completed, which causes the $e_0$ generated by truncation to accumulate until it is reshared. The reason why the triangle (green) line ultimately reaches the highest accuracy is that even though the error $e_0$ is accumulated, it becomes negligible as the precision increases.

Lastly, we would like to point out that using the ``truncate-then-multiply'' approach may incur a loss of precision compared to the ``multiply-then-truncate'' method. However, this loss of precision is much smaller than the error introduced by $e_1$. Based on our experimental results, we observed that the overall end-to-end inference accuracy is hardly affected, and therefore, we refrain from further elaborating on this error in this paper.

\section{Impact of $e_1$ on Efficiency in PPML and Propoesed Solutions} \label{sec:drelu}
In this section, we will delve into how $e_1$ becomes a factor leading to performance bottlenecks in certain protocols due to the constraints imposed by parameter choices. Taking Bicoptor~\cite{ZWC+-23} as an example, the work selected $\ell=64$ to ensure enough slack, the one-pass dominating communication cost was about $\ell_x\cdot\ell = 31\cdot64\approx2,000$ bits. If there is no need to ensure a sufficiently large slack, $\ell$ could be chosen as $31$, and the communication cost would be reduced by half. To address this problem, we propose a non-interactive deterministic truncation protocol. Note that the new truncation protocol is designed to replace the probabilistic truncation protocol used in the nonlinear layer specifically. It cannot be directly applied to solve the problems caused by $e_1$ in linear layers.

We present this new non-interactive deterministic truncation protocol in Sect.~\ref{sub:newtrc}. In Sect.~\ref{sub:cmp}, we will review the core principle of using truncation protocol for sign determination in Bicoptor~\cite{ZWC+-23}, and theoretically demonstrate that the new truncation protocol is also applicable to this principle. During the process of designing a new three-party DReLU protocol based on this principle, we identify some potential security and correctness issues, which we will address in Sect.~\ref{sub:modswitch} by presenting a proposed solution. Finally, combining the aforementioned content with some optimizations, we will showcase our new Bicoptor 2.0 DReLU protocol in Sect.~\ref{sub:drelu}.

\subsection{A Non-interactive Deterministic Truncation Protocol} \label{sub:newtrc}
By Theorem~\ref{thm:cut}, we know that the error term $e_1$ of the general cut function is always $\mathsf{LT}(\cdot,\cdot)\cdot\mathsf{cut}(2^\ell,k)$. The question is how could we eliminate this item. A natural way is to modulo the cut function by $2^{l-k}$, since $\mathsf{cut}(2^\ell,k) = 2^{l-k}$. Alg.~\ref{alg:newtrc} presents our new non-interactive deterministic truncation protocol. 
\begin{algorithm}[ht]
  \textbf{Input}: shares of $x\in[0,2^{\ell_x})\bigcup(2^\ell-2^{\ell_x},2^\ell)$ in $\mathbb{Z}_{2^\ell}$, number of bits to be truncated $k$\\
  \textbf{Output}: shares of $\mathsf{trc}(x,k)$ in $\mathbb{Z}_{2^{\ell-k}}$
      \begin{algorithmic}[1]
        \State $P_0$ sets $[\overline{\mathsf{trc}}(x,k)]_0:=\mathsf{cut}([x]_0,k)\ \text{mod}\ 2^{\ell-k}$.
        \State $P_1$ sets $[\overline{\mathsf{trc}}(x,k)]_1:=2^\ell - \mathsf{cut}(2^\ell - [x]_1,k)\ \text{mod}\ 2^{\ell-k}$.
      \end{algorithmic}
    \caption{The Non-interactive Deterministic Truncation Protocol.}
    \label{alg:newtrc}
\end{algorithm}

We reuse the example used in a previous section to demonstrate how $e_1$ can lead to serious errors, to better illustrate how the new truncation protocol (Alg.~\ref{alg:newtrc}) eliminates $e_1$. Recall that, in the previous example, $x = 0100\ 1011$, $R = 1110\ 0000$, $x + R\ \text{mod}\ 2^8 = 0010 1011 < x$ and hence, $e_1$ occurs. 
\begin{align*}
	& x = 0100\ 1011, R = 1110\ 0000, \ell = 8, k = 4,\\
  & [x]_0 = x + R\ \text{mod}\ 2^8 = 0010\ 1011, \\
  & [x]_1 = -R\ \text{mod}\ 2^8 = 0010\ 0000 \\
&\overline{\mathsf{trc}}(x,4) = [\overline{\mathsf{trc}}(x,4)]_0 + [\overline{\mathsf{trc}}(x,4)]_1\ \text{mod}\ 2^{8-4} \\
= &(\mathsf{cut}([x]_0,4)\ \text{mod}\ 2^{8-4} - \mathsf{cut}(-[x]_1,4)\ \text{mod}\ 2^{8-4})\\
  &\ \text{mod}\ 2^{8-4}\\
= & 0010 - 1110\ \text{mod}\ 2^{8-4} = 0100
\end{align*}


We further notice that all truncation protocols discovered so far are focusing on cutting the last few bits of input. Alg.~\ref{alg:newtrc2} presents an extension of Alg.~\ref{alg:newtrc} which allows us to obtain the middle few bits of input, i.e., cut off the last few bits and the first few bits. Such truncation protocol remains non-interactive and deterministic. 
\begin{algorithm}[ht]
  \textbf{Input}: shares of $x\in[0,2^{\ell_x})\bigcup(2^\ell-2^{\ell_x},2^\ell)$ in $\mathbb{Z}_{2^\ell}$, first $k_1$ bits to be truncated and last $k_2$ bits to be truncated\\
  \textbf{Output}: shares of $\mathsf{trc}(x,k)$ in $\mathbb{Z}_{2^{\ell-k_1-k_2}}$
      \begin{algorithmic}[1]
        \State $P_0$ sets $$[\overline{\mathsf{trc}}(x,k_1,k_2)]_0:=\mathsf{cut}([x]_0,k_1,k_2)\ \text{mod}\ 2^{\ell-k_1-k_2}.$$
        \State $P_1$ sets $$[\overline{\mathsf{trc}}(x,k_1,k_2)]_1:=2^\ell - \mathsf{cut}(2^\ell - [x]_1,k_1,k_2)\ \text{mod}\ 2^{\ell-k_1-k_2}.$$
      \end{algorithmic}
    \caption{The More General Non-interactive Deterministic Truncation Protocol.}
    \label{alg:newtrc2}
\end{algorithm}

An example of Alg.~\ref{alg:newtrc2} is given below:
\begin{align*}
	& x = 0100\ 1011, R = 1110\ 0000, \ell = 8, k_1 = 4, k_2 = 1\\
  & [x]_0 = x + R\ \text{mod}\ 2^8 = 0010\ 1011, \\
  & [x]_1 = -R\ \text{mod}\ 2^8 = 0010\ 0000 \\
	& \overline{\mathsf{trc}}(x,4,1) =[\overline{\mathsf{trc}}(x,4,1)]_0 + [\overline{\mathsf{trc}}(x,4,1)]_1\ \text{mod}\ 2^{8-4-1}\\
   = & ((\mathsf{cut}([x]_0,4,1)\ \text{mod}\ 2^{8-4-1} \\
   &- \mathsf{cut}(-[x]_1,4,1)\ \text{mod}\ 2^{8-4-1}))\ \text{mod}\ 2^{8-4-1}\\
   =& 010 - 110\ \text{mod}\ 2^{8-4-1} = 100
\end{align*}
The correctness of Alg.~\ref{alg:newtrc} and Alg.~\ref{alg:newtrc2} can be proven by the following theorem~\ref{thm:newcut2}, for details, please refer to Appendix~\ref{app:sub:proof:thm:newcut2}.
\begin{theorem} \label{thm:newcut2}
  For $\alpha,\beta\in\mathbb{Z}_{2^\ell}$ and $\mathsf{bit} := \{0,1\}$,
  \begin{itemize}[leftmargin=*]
    \item $\mathsf{cut}(\alpha+\beta,k_1,k_2)\ \text{mod}\ 2^{\ell-k_1-k_2}=\mathsf{cut}(\alpha,k_1,k_2)+\mathsf{cut}(\beta,k_1,k_2)+\mathsf{bit}\ \text{mod}\ 2^{\ell-k_1-k_2}$;
    \item $\mathsf{cut}(\alpha-\beta,k_1,k_2)\ \text{mod}\ 2^{\ell-k_1-k_2}=\mathsf{cut}(\alpha,k_1,k_2)-\mathsf{cut}(\beta,k_1,k_2)-\mathsf{bit}\ \text{mod}\ 2^{\ell-k_1-k_2}$;
  \end{itemize}
\end{theorem}

\subsection{The Principle of Sign Determination} \label{sub:cmp}
We aim to replace the truncation protocol proposed by SecureML~\cite{MZ17} in Bicoptor~\cite{ZWC+-23} with our new protocol to address the issues caused by the probabilistic truncation protocol, and thus obtain a more efficient DReLU protocol. However, replacing the main sub-protocols in such a complex DReLU protocol is not a trivial task, and it needs to be further proven whether the new DReLU protocol with the replaced sub-protocol is still effective. In this sub-section, we will explain the core principles of the DReLU protocol in Bicoptor~\cite{ZWC+-23} and provide the core theory, lemmas, and proofs for the DReLU protocol that are applicable to the new truncation protocol in Sect.~\ref{sub:newtrc}. 

We recall that the key idea is to determine the output of $\mathsf{trc}(x,\lambda)$ or $\mathsf{trc}(x,\lambda -1)$, where $\lambda$ is the effective bit length of $\xi$ and $\xi$ is the ``absolute value'' of $x$. In the following lemmas and theorem, we use $\overline{\mathsf{trc}}(x,\lambda-1)$ instead of $\overline{\mathsf{trc}}(x,\lambda-1,k)$ for $k\leq\ell-\lambda-1$ because when truncating the effective bits, the outputs of are the same, either 0 or 1, regardless of the $k$. By Lemma~\ref{lmm:pattern3}, we have $\overline{\mathsf{trc}}(x,\lambda)$ or $\overline{\mathsf{trc}}(x,\lambda -1)$ is $1$ for positive input and $2^\ell-1$ for negative input. Due to $\lambda$ being unknown, we need to perform $\ell_x$ times of truncation and output an array of outcomes. Lemma~\ref{lmm:pattern4} proves the necessary existence of $1$ or $2^\ell-1$ in this array, and Lemma~\ref{lmm:pattern5} shows a specific pattern of the array. Hence, we can simply check the existence of $1$ or $2^\ell-1$ to determine the sign. i.e., Theorem~\ref{thm:pattern0}. The proofs of Lemma~\ref{lmm:pattern3},~\ref{lmm:pattern4} and~\ref{lmm:pattern5} could be found in Appendix~\ref{app:sub:proof:thm:pattern0}. All these lemmas imply the following theorem: Theorem~\ref{thm:pattern0}.

\begin{theorem} \label{thm:pattern0}
  For an input $x\in\mathbb{Z}_{2^\ell}$ with precision of $\ell_x$, let $\xi:=x$ if $x$ is positive, and let $\xi:=2^\ell-x\ \text{mod}\ 2^\ell$ if $x$ is negative. The binary form $\xi$ is defined as $\{\xi_{\ell_x-1},\xi_{\ell_x-2},\cdots,\xi_{1},\xi_{0}\}$, where $\xi_i$ denotes the $i$-th bit of $\xi$. $\lambda$ is the effective bit length of $\xi$, i.e., $\xi_{\lambda-1} = 1$ and $\lambda + 1 < \ell$. Set $\xi:=\xi^\prime\cdot 2^k + \xi^{\prime\prime}$, where $\xi^\prime\in[0, 2^{\ell_x-k})$ and $\xi^{\prime\prime}\in[0, 2^k)$. We have that for any $\hat{\ell} \geq \lambda$, the following results hold:
\begin{itemize} [leftmargin=*]
  \item For a positive $x$, there exists positive numbers $\lambda^\prime$ and $\lambda^{\prime\prime}$ ($\lambda^\prime \leq \lambda^{\prime\prime} \leq \ell_x$) satisfying $\overline{\mathsf{trc}}(\xi, j) = 1$ for $\lambda^\prime \leq j \leq \lambda^{\prime\prime}$, and $\overline{\mathsf{trc}}(\xi,j) = 0$ for $j > \lambda^{\prime\prime}$. 
  \item For a negative $x$, there exists positive numbers $\lambda^\prime$ and $\lambda^{\prime\prime}$ ($\lambda^\prime \leq \lambda^{\prime\prime} \leq \ell_x$) satisfying $\overline{\mathsf{trc}}(2^\ell - \xi, j) = 2^\ell - 1$ for $\lambda^\prime \leq j \leq \lambda^{\prime\prime}$, and $\overline{\mathsf{trc}}(2^\ell - \xi,j) = 0$ for $j > \lambda^{\prime\prime}$. 
\end{itemize}     
\end{theorem}

To help us better understand Theorem~\ref{thm:pattern0}, we provide the following example using the extended truncation protocol Alg.~\ref{alg:newtrc2} in Tab.~\ref{tab:dreluexp} for both positive and negative inputs. In this example, $\ell=64$, $\ell_x=7$, and $\lambda=5$. When $x = 0...0010110$ is positive, $\overline{\mathsf{trc}}(x,4,\ell-\ell_x-4)=0000001$, and $\overline{\mathsf{trc}}(x,k,\ell-\ell_x-k)$ for $k>4$ would be $0$ if no $e_0$ occurs. Similarly, for negative $x = 2^{64} - 0...0010110=1...1110110$, $\overline{\mathsf{trc}}(x,4,\ell-\ell_x-4)=2^{64} - 1\ \text{mod}\ 2^7=1111111$, and $\overline{\mathsf{trc}}(x,k,\ell-\ell_x-k)$ for $k>4$ would be $0$ if no $e_0$ occurs.

\begin{table}[t!]
  \caption{An example of the sign determination.}
  \label{tab:dreluexp}
  \centering
  \linespread{1.25}
  \footnotesize
  \begin{tabular*}{8.5 cm}{@{\extracolsep{\fill}} l p{1.4cm} p{0.3cm} p{2.1cm} p{0.3cm}}
    \multirow{2}*{$\ell=64,\ell_x=7,\lambda=5$} & \multicolumn{2}{l}{\multirow{2}*{$x=0...0010110$}} & \multicolumn{2}{c}{$x = 2^{64} - 0...0010110$}\\
    &   & & \multicolumn{2}{c}{$= 1...1110110$}\\
    \toprule
    \textbf{Operation} & \textbf{Value} & \textbf{$e_0$} & \textbf{Value} & \textbf{$e_0$} \\
    \hline
    $\overline{\mathsf{trc}}(x,0,\ell-\ell_x-0)$ & $0010110$   & $0$  & $1101010$ & $0$ \\          
    $\overline{\mathsf{trc}}(x,1,\ell-\ell_x-1)$ & $0001011$   & $0$  & $1110101$ & $0$ \\          
    $\overline{\mathsf{trc}}(x,2,\ell-\ell_x-2)$ & $0000110$   & $+1$ & $1111010$ & $-1$\\           
    $\overline{\mathsf{trc}}(x,3,\ell-\ell_x-3)$ & $0000010$   & $0$  & $1111110$ & $0$ \\ 
    $\overline{\mathsf{trc}}(x,4,\ell-\ell_x-4)$ & $0000001$   & $0$  & $1111111$ & $0$ \\
    $\overline{\mathsf{trc}}(x,5,\ell-\ell_x-5)$ & $0000000$   & $0$  & $0000000$ & $0$ \\
    $\overline{\mathsf{trc}}(x,6,\ell-\ell_x-6)$ & $0000000$   & $0$  & $0000000$ & $0$ \\
    $\overline{\mathsf{trc}}(x,7,\ell-\ell_x-7)$ & $0000000$   & $0$  & $0000000$ & $0$ \\
    \bottomrule
  \end{tabular*}
\end{table}

\subsection{Modulo-switch Protocol} \label{sub:modswitch}
In this subsection, we present the modulo-switch protocol and explain how it enhances our sign determination principle. From a theoretical perspective, we find that using the new truncation protocol is feasible to determine the sign of a number, again, sign determination is equivalent to DReLU. However, when it comes to a practically applicable DReLU protocol, replacing the truncation protocol with the new one still presents security and correctness issues. Therefore, we propose the modulo-switch protocol, which completes our new Bicoptor 2.0 DReLU protocol. 

From the security point of view, while working in $\mathbb{Z}_{2^{\ell^\prime}}$ (e.g., $\ell^\prime=\ell-k_1-k_2$), the parity of a number remains after masking, more specifically, if the product of two numbers is odd, this means both two numbers are odd. Leaking the parity of data could cause other more serious issues. By switching the ring from $\mathbb{Z}_{2^{\ell^\prime}}$ to $\mathbb{Z}_{p}$ for a prime $p$ could prevent this issue. From the correctness point of view, we want non-zero outputs remain non-zero after masking and zero outputs remain zero. However, while working in $\mathbb{Z}_{2^{\ell^\prime}}$, one non-zero output could possibly become zero after multiplying a random number. For example, $16\cdot16\ \text{mod}\ 2^8 = 0$. We present the modulo-switch protocol in Alg.~\ref{alg:modswitch}:

\begin{algorithm}[ht]
  \textbf{Input}: shares of $x$ in $\mathbb{Z}_{2^{\ell^\prime}}$\\
  \textbf{Output}: shares of $x$ in $\mathbb{Z}_p$, where $\log_2p=\ell^\prime+1$.
    \begin{algorithmic}[1]
      \State $P_0$ sets $[x]_0:=2^{\ell^\prime}\ \text{mod}\ p$ if $[x]_0=0$,
      otherwise sets $[x]_0:=[x]_0\ \text{mod}\ p$.
      \State $P_1$ sets $[x]_1:= p + [x]_1 - 2^{\ell^\prime}\ \text{mod}\ p$.
    \end{algorithmic}
  \caption{Modulo-switch Protocol.}
  \label{alg:modswitch}
\end{algorithm}

It is easy to argue the correctness of Alg.~\ref{alg:modswitch}. The aim of Alg.~\ref{alg:modswitch} is to ensure the output elements in $\mathbb{Z}_p$ are zero if and only if the input elements in $\mathbb{Z}_{2^{\ell^\prime}}$ are zero. Considering $x=[x]_0+[x]_1\ \text{mod}\ 2^{\ell^\prime}=0$:
\begin{itemize}[leftmargin=*]
  \item If $[x]_0\ \text{mod}\ 2^{\ell^\prime}=0$, then $[x]_1\ \text{mod}\ 2^{\ell^\prime}=0$. $P_0$ sets $[x]_0:=2^{\ell^\prime}\ \text{mod}\ p$ and $P_1$ sets $[x]_1:=p - 2^{\ell^\prime}\ \text{mod}\ p$. The output is $[x]_0 + [x]_1 = p = 0\ \text{mod}\ p$.
  \item If $[x]_0\ \text{mod}\ 2^{\ell^\prime}\neq0$, then $[x]_1 = 2^{\ell^\prime} - [x]_0\ \text{mod}\ 2^{\ell^\prime}$. $P_0$ sets $[x]_0:=[x]_0\ \text{mod}\ p$ and $P_1$ sets $[x]_1:=p + 2^{\ell^\prime} - [x]_0 - 2^{\ell^\prime} = p - [x]_0\ \text{mod}\ p$. The output is $[x]_0 + [x]_1 = p = 0\ \text{mod}\ p$.
\end{itemize}    

\subsection{DReLU Protocol} \label{sub:drelu}
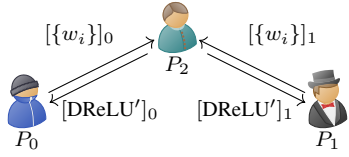
\begin{figure}[ht!]
  \centering
  \begin{tikzpicture}[x=1cm,y=1cm,cap=round,align=center,
      fact/.style={rectangle, draw, rounded corners=1mm, fill=white, drop shadow,
            text centered, anchor=center, text=black},growth parent anchor=center,
      fact2/.style={rectangle, draw, rounded corners=1mm, fill=white,
            text centered, anchor=center, text=black},growth parent anchor=center]
  
  
      \node (u2) [charlie, minimum size=0.5cm] at (-5,-0.5) {};
      \node at (-5,-1) [draw=none, anchor=center] {\footnotesize $P_2$};
      \node (u0) [criminal, minimum size=0.5cm] at (-7,-1.5) {};
      \node at (-7,-2) [draw=none, anchor=center] {\footnotesize $P_0$};
      \node (u1) [groom, minimum size=0.5cm] at (-3,-1.5) {};
      \node at (-3,-2) [draw=none, anchor=center] {\footnotesize $P_1$};
  
      \draw [<-] (u2) -- (u0);
      \draw [<-] (u2) -- (u1);

      \node at (-6.3,-0.6) [draw=none, anchor=center] {\footnotesize $[\{w_i\}]_0$};
      \node at (-3.6,-0.6) [draw=none, anchor=center] {\footnotesize $[\{w_i\}]_1$};

      \draw [<-] (-6.7,-1.5) -- (-5.4,-0.85);
      \draw [<-] (-3.3,-1.5) -- (-4.6,-0.85);

      \node at (-5.9,-1.6) [draw=none, anchor=center] {\footnotesize $[\text{DReLU}^\prime]_0$};
      \node at (-4.1,-1.6) [draw=none, anchor=center] {\footnotesize $[\text{DReLU}^\prime]_1$};
    
  \end{tikzpicture}
  \caption{The Overview of Bicoptor 2.0 DReLU Protocol.}
  \label{fig:dreluoverview}
\end{figure}

With both deterministic truncation protocol and modulo-switch protocol, we now present our new DReLU protocol in Alg.~\ref{alg:drelu} and we will provide a step-by-step explanation. The overall system setting is that $P_0$ and $P_1$ locally compute arrays $[\{w_i\}]_0$ and $[\{w_i\}]_1$ then sent them to $P_2$. $P_2$ as an assisting party, helps reconstruct $\{w_i\}$ and returns an intermediate DReLU result to $P_0$ and $P_1$, who compute the final DReLU result. A more intuitive system architecture diagram is shown in Fig.~\ref{fig:dreluoverview}. 

\begin{algorithm}[ht]
  \textbf{Setting}: $\ell$, $\ell_x$, and $p$. $P_0$ and $P_1$ share $\mathsf{seed}_{01}$.\\
  \textbf{Input}: shares of $x$\\
  \textbf{Output}: shares of $\text{DReLU}(x)$
  \begin{algorithmic}[1]
    \Statex // \textit{$P_0$,$P_1$ initialize.}
    \State $P_0$ and $P_1$ generate a random bit $t$ using $\mathsf{seed}_{01}$. \label{alg:drelu:prep:step1}
    \State $P_0$ and $P_1$ compute $[x]:=(-1)^t\cdot[x]\ \text{mod}\ 2^\ell$. \label{alg:drelu:prep:step2}
    \State $P_0$ and $P_1$ compute \label{alg:drelu:prep:step3}
    $$[u_i]:=[\overline{\mathsf{trc}}(x,i,\ell-\ell_x-i)]\ \text{mod}\ 2^{\ell_x}, \forall i\in[0,\ell_x].$$
    \State $P_0$ and $P_1$ compute \label{alg:drelu:prep:step4}
    $$[v_i]:=[u_i] + [u_{i+1}] - 1\ \text{mod}\ 2^{\ell_x}, \forall i\in[0,\ell_x-1]$$
    and $[v_{\ell_x}]:=[u_{\ell_x}] - 1\ \text{mod}\ 2^{\ell_x}$.
    \State $P_0$ and $P_1$ run Alg.~\ref{alg:modswitch} to switch the ring from $\mathbb{Z}_{2^{\ell_x}}$ to $\mathbb{Z}_{p}$ for $[v_i]$. \label{alg:drelu:prep:step5}
    \State $P_0$ and $P_1$ use $\mathsf{seed}_{01}$ to shuffle $[\{v_i\}]:=\Pi([\{v_i\}])$. \label{alg:drelu:prep:step6}
      \State $P_0$ and $P_1$ generate $\ell_x + 1$ numbers of random   \label{alg:drelu:prep:step7}
      \Statex $\{r_i\}$ using $\mathsf{seed}_{01}$, where $r_i\in\mathbb{Z}_p^*,\mathbb{Z}_p^*:=\mathbb{Z}_p/\{0\}$.
      \Statex Masking by performing $[\{w_i\}]:=[\{v_i\cdot r_i\}]\ \text{mod}\ p$.
      \State $P_0$ and $P_1$ reshare and send $[\{w_i\}]$ to $P_2$. \label{alg:drelu:prep:step8}
      \Statex // \textit{$P_2$ processes.}
      \State $P_2$ reconstructs $\{w_i\}$, and sets $\text{DReLU}(x)^\prime:=1$ if \label{alg:drelu:prep:step9}
      \Statex there is 0 in $\{w_i\}$, otherwise sets $\text{DReLU}(x)^\prime:=0$.
      \State $P_2$ responds $[\text{DReLU}(x)^\prime]\in\mathbb{Z}_{2^{\ell}}$ to $P_0$ and $P_1$.\label{alg:drelu:prep:step10}
      \Statex // \textit{$P_0$ and $P_1$ finalize.}
      \State $P_0$ and $P_1$ compute $[\text{DReLU}(x)]=[\text{DReLU}(x)^{\prime} \oplus t]$ \label{alg:drelu:prep:step11}
      \Statex $=t + (1-2t)\cdot[(\text{DReLU}(x)^{\prime}]\ \text{mod}\ 2^\ell$.
    \end{algorithmic}
  \caption{Bicoptor 2.0 DReLU Protocol.}
  \label{alg:drelu}
\end{algorithm}

\noindent \textbf{Step-by-Step Explanation} 


\begin{itemize} [leftmargin=*]
  \item In step 1-2, $P_0$ and $P_1$ blind their input shares $[x]_0$ and $[x]_1$ using random bit $t$ generated from preshared $\mathsf{seed}_{01}$, i.e., the sign of the input has been randomly flipped.
  \item In step 3, as mentioned in Sect.~\ref{sub:cmp}, $\lambda$ is unknown and hence, $P_0$ and $P_1$ repeatedly compute $\ell_x$ number of times of truncations and obtain $[\{u_i\}]_0$ and $[\{u_i\}]_1$. Note that $[\{u_i\}]_0$ and $[\{u_i\}]_1$ are elements in $\mathbb{Z}_{2^{\ell_x}}$ instead of that in $\mathbb{Z}_{2^{\ell}}$ proposed in Bicoptor~\cite{ZWC+-23}, due to the usage of our new truncation protocol. We remove the $u_{*}$ term which was originally included in Bicoptor~\cite{ZWC+-23}. The purpose of $u_{*}$ was to help determine $\text{DReLU}(0)$, but our ultimate goal is to improve the performance of ReLU and thus the end-to-end inference performance. Since $\text{ReLU}(x) = \text{DReLU}(x)\cdot x$, when $x=0$, $\text{ReLU}(0) = \text{DReLU}(0)\cdot0$, regardless of the result of $\text{DReLU}(0)$, the final result of $\text{ReLU}(0)$ is always $0$, therefore removng the $u_{*}$ term will not effect the E2E PPML inference.
  \item In step 4, $P_0$ and $P_1$ locally perform adjacent pairwise addition on $[u_i]$, i.e., $[u_i] + [u_{i+1}]\ \forall i\in[0,\ell_x-1]$. Recall that the original summation proposed in Bicoptor~\cite{ZWC+-23} was recursive summation, i.e., $\Sigma_{k=i}^{\ell_x}[u_k],\ \forall i\in[0,\ell_x]$. This modification was made to enable better parallelization of the protocol and to reduce computational overhead.
  \item In step 5, $P_0$ and $P_1$ run Alg.~\ref{alg:modswitch} to switch the ring from $\mathbb{Z}_{2^{\ell_x}}$ to $\mathbb{Z}_{p}$ for $[v_i]$.
  \item In step 6-7, shuffling and masking are performed to avoid information leakage of the input.~\footnote{More explanation can be found in Bicoptor~\cite[Sect. 3.3]{ZWC+-23}.}
  \item In step 8, $P_0$ and $P_1$ reshare and send $[\{w_i\}]_0$ and $[\{w_i\}]_1$ to $P_2$, note that the one-pass dominating communication overhead is now reduced from $\ell_x\cdot\ell$ to $(\ell_x + 1)\cdot(\ell_x + 1)$ bits. In the case of $\ell = 64$ and $\ell_x = 31$, the communication overhead has been reduced by half. 
  \item In step 9-10, $P_2$ reconstructs $\{w_i\}$ and outputs the intermediate result $[\text{DReLU}(x)^\prime]\in\mathbb{Z}_{2^\ell}$. $P_2$ then sends $[\text{DReLU}(x)^\prime]_0$ and $[\text{DReLU}(x)^\prime]_1$ back to $P_0$ and $P_1$, respectively. 
  \item Finally, in step 11, $P_0$ and $P_1$ unblind $[\text{DReLU}(x)^\prime]$ using the random bit $t$ and construct $[\text{DReLU}(x)]$ locally, i.e., $[\text{DReLU}(x)]=[\text{DReLU}(x)^{\prime} \oplus t] =t + (1-2t)\cdot[(\text{DReLU}(x)^{\prime}]\ \text{mod}\ 2^\ell$. 
\end{itemize}

\noindent \textbf{From DReLU to ReLU.} After obtaining the $\text{DReLU}(x)$, computing $\text{ReLU}(x) = x\cdot \text{DReLU}(x)$ becomes relatively straightforward. A common approach is to use a Beaver triple~\cite{B91} and work~\cite{ZWC+-23} introduces a method to generate the shares of a triple using pre-shared seeds ($\mathsf{seed}_{02}$ and $\mathsf{seed}_{12}$). The more detailed ReLU protocol is presented in Appendix~\ref{app:relu}.

\noindent \textbf{From UBL to RSS.} We want to emphasize that the new DReLU and ReLU protocols both work with replicated secret sharing. More details can be found in Appendix~\ref{app:rss}.

\section{Performance Evaluation} \label{sec:evaluate}
In this section, we first introduce how we discover that in the DReLU layer, only using a portion of the data bits,~\footnote{For an 8-bit input, $01001011$, we are only using the top five bits as the DReLU input, i.e., $01001$.} which we call them the key bits, has a negligible impact on the inference accuracy. We also present experimental results on which bits are the key-bits for different models in Sect.~\ref{sub:keybit}. Next, we compare the performance of our DReLU/ReLU protocols based on these key bits with the performance of the original DReLU/ReLU protocol used in P-Falcon~\cite{WWP22,piranha-code} in Sect.~\ref{sub:unittest}. Finally, we demonstrate the E2E PPML inference performance of the overall system through experimental results in Sect.~\ref{sub:e2etest}.

We use three cloud server nodes to simulate three parties, each node with the following configuration: two Intel(R) Xeon(R) E5-2690 v4 @ 2.60GHz CPUs, 64 GiB memory, and one independent Nvidia Tesla P100 GPU. Each node uses Ubuntu 16.04.7 and CUDA 11.4. We also simulate three different network environments: LAN1, LAN2, and WAN corresponded to 5Gbps/1Gbps/100Mbps bandwidth and 0.2ms/0.6ms/40ms round trip latency, respectively. Finally, we used four models in our experiments, denoted as CIFAR10\_AlexNet, CIFAR10\_VGG16, Tiny\_AlexNet, and Tiny\_VGG16 with parameter $\ell=64$.

\begin{table}[t!]
  \caption{Aiming for the highest accuracy: exploring different combinations of $\ell_x^\mathsf{int}$ and $\ell_x^\mathsf{frac}$ for improved inference in various models.}
  \label{tab:accuracy2}
  \centering
  \linespread{1.25}
  \footnotesize
  \begin{tabular*}{8.5cm}{@{\extracolsep{\fill}} p{0.9cm} l p{1cm} p{1cm} p{1cm}}
	  \toprule
	  \textbf{Model} & \textbf{DReLU Precision} & \textbf{PPML Acc.} & \textbf{Plaintext Acc.} & \textbf{Acc. Lose}\\
	  \hline
	  \multirow{4}{1cm}{CIFAR10\_\\AlexNet} 
	  & $\ell_x=\ell_x^\mathsf{int}+\ell_x^\mathsf{frac}=4+3$ & 62.69\% & \multirow{4}*{69.63\%} & 6.94\% \\
	  & \textbf{$\ell_x=\ell_x^\mathsf{int}+\ell_x^\mathsf{frac}=5+2$} & \textbf{69.28\%} & & \textbf{0.35\%} \\
	  & $\ell_x=\ell_x^\mathsf{int}+\ell_x^\mathsf{frac}=6+1$ & 68.67\% & & 0.96\%\\
	  & $\ell_x=\ell_x^\mathsf{int}+\ell_x^\mathsf{frac}=7+0$ & 67.67\% & & 1.96\%\\
	  \hline
	  \multirow{4}{1cm}{Tiny\_\\AlexNet} 
	  & $\ell_x=\ell_x^\mathsf{int}+\ell_x^\mathsf{frac}=4+3$ & 3.26\% & \multirow{4}*{26.39\%} & 23.13\% \\
	  & $\ell_x=\ell_x^\mathsf{int}+\ell_x^\mathsf{frac}=5+2$ & 21.12\% & & 5.27\% \\
	  & $\ell_x=\ell_x^\mathsf{int}+\ell_x^\mathsf{frac}=6+1$ & 25.80\% & & 0.59\% \\
	  & \textbf{$\ell_x=\ell_x^\mathsf{int}+\ell_x^\mathsf{frac}=7+0$} & \textbf{25.89\%} & & \textbf{0.50\%} \\
	  \hline
	  \multirow{4}{1cm}{CIFAR10\_\\VGG16}
	  & $\ell_x=\ell_x^\mathsf{int}+\ell_x^\mathsf{frac}=4+3$ & 88.36\% & \multirow{4}*{88.31\%} & No loss \\
	  & \textbf{$\ell_x=\ell_x^\mathsf{int}+\ell_x^\mathsf{frac}=5+2$} & \textbf{88.43\%} & & \textbf{No loss} \\
	  & $\ell_x=\ell_x^\mathsf{int}+\ell_x^\mathsf{frac}=6+1$ & 88.42\% & & No loss \\
	  & $\ell_x=\ell_x^\mathsf{int}+\ell_x^\mathsf{frac}=7+0$ & 88.40\% & & No loss \\
	  \hline
	  \multirow{4}{1cm}{Tiny\_\\VGG16}
	  & $\ell_x=\ell_x^\mathsf{int}+\ell_x^\mathsf{frac}=4+3$ & 0.78\% & \multirow{4}*{54.89\%} & 54.11\% \\
	  & $\ell_x=\ell_x^\mathsf{int}+\ell_x^\mathsf{frac}=5+2$ & 21.95\% & & 32.94\% \\
	  & $\ell_x=\ell_x^\mathsf{int}+\ell_x^\mathsf{frac}=6+1$ & 54.55\% & & 0.34\% \\
	  & \textbf{$\ell_x=\ell_x^\mathsf{int}+\ell_x^\mathsf{frac}=7+0$} & \textbf{54.94\%} & & \textbf{No loss} \\
	  \bottomrule
  \end{tabular*}
\end{table}

\subsection{Ultimate Performance: Key Bits} \label{sub:keybit}
After optimizing the DReLU protocol, our theoretical communication cost has already reached $(\ell_x+1)\cdot(\ell_x+1)$ bits, where $\ell_x$ is generally 31 for $\ell = 64$. However, we do not stop there. We also explore the possibility of reducing $\ell_x$ further to improve performance. Recall that, the ReLU function preserves the original value for positive numbers and sets negative numbers to $0$. We speculate that for a float like $5.123456$, if we only take $5.12$ as the ReLU input, it would not have a significant impact on the inference results. We conduct extensive experiments, and the results are shown in Tab.~\ref{tab:accuracy2}. We find that using $\ell_x=\ell_x^\mathsf{int}+\ell_x^\mathsf{frac}=5+2$ in CIFAR10\_AlexNet or $\ell_x=\ell_x^\mathsf{int}+\ell_x^\mathsf{frac}=7 + 0$ in Tiny\_AlexNet only causes a 0.35\% or 0.50\% decrease in inference accuracy, respectively, while performance could be significantly improved. We believe that this trade-off is highly efficient. We would like to kindly remind that using key bits is only applied in the DReLU protocol, when computing $\text{ReLU}(x) = x\cdot \text{DReLU}(x)$. The coefficient $x$ used in this multiplication should still utilize all precision bits.

The ``key bits'' optimization is a highly versatile trick that is applicable to the majority of sign determination protocols. In Tab.~\ref{tab:compare}(lower section), we provide the theoretical communication costs for other works when they incorporate our ``key bits'' optimization. However, its implementation is not straightforward due to the limitation of existing truncation protocols, which cannot selectively extract a portion of the data. In order to provide a better understanding of our work, I divided the ``key bits'' optimization into two stages for discussion, and we apply this optimization to compare its performance with Bicoptor~\cite{ZWC+-23}.

Let's consider an 64-bit input $x = 0...1001...1011$ with $\ell_x = 32$, and select the first 4 bits of the effective bits of x, namely 1001 as the key bits. First stage: $x_\mathsf{keyBits} = 0...1001$ (64 bits) with communication overhead reduced down to $\ell\cdot\ell_x = 64\cdot4 = 256$. Second stage: $x_\mathsf{keyBits} = 1001$ (4 bits) with communication overhead reduced down to $5\cdot4 = 20$.

The first stage of the ``key bits'' optimization is highly versatile but with limited improvement in performance. For instance, in the lower section of Tab.~\ref{tab:compare}, we observe that the communication cost of Bicoptor~\cite{ZWC+-23} is only reduced from $2048$ bits to $512$ bits. However, even though it is widely applicable, we are the first to implement the utilization of a specific portion of the data in the MPC setting for calculating DReLU. We conducted extensive experiments to determine the optimal selection of key bits, thus finding an optimal balance between accuracy and performance. The second stage of the ``key bits'' optimization can further enhance the performance of the DReLU protocol. However, it requires the utilization of our deterministic truncation protocol. With both stages of the ``key bits'' optimization, the communication cost of Bicoptor~\cite{ZWC+-23} is now reduced from $2048$ bits to $64$ bits (Tab.~\ref{tab:compare}).


\subsection{Bicoptor 2.0 DReLU/ReLU Unit Experiments}  \label{sub:unittest}
We implement our DReLU and ReLU protocols and evaluate their performance with different batch sizes in various network environments. We compare our results with the DReLU/ReLU protocols in P-Falcon~\cite{WWP22,piranha-code}, which are shown in Tab.~\ref{tab:drelu} and Tab.~\ref{tab:relu}, respectively. We observe that in the LAN1 network environment, our DReLU protocol achieves over 10x speedup for batch size = $10^5$, while ReLU achieves approximately 6-7x speedup. A more intuitive visualization of the results can be seen in Fig.~\ref{fig:relu}.

\begin{figure*}[t!] 
  \centering 
      \begin{tikzpicture}
      \footnotesize
      \begin{axis}[
        height=3cm, width= 4.4cm, axis lines=middle,
        axis line style={->}, 
        ymin=0, ymax=20000000,
        xmode=log,
        x label style={at={(axis description cs:0.5,-0.25)},anchor=north},
        y label style={at={(axis description cs:-0.2,0.5)},rotate=90,anchor=south},
        title={LAN1},
        title style={at={(axis description cs:0.5,0.9)},anchor=north},
          xlabel=Batch size,
          ylabel=Throughput rate (ops),
          legend style={at={(-0.5,0.5)},anchor=east}, legend columns=1]
      
      \addplot[smooth,mark=-,gray] plot coordinates {
          (1,163.01) (1000,151815.56) (10000,874951.88) (100000,1661350.35) (1000000,1666433.37)
      };
      \addlegendentry{P-Falcon DReLU~\cite{WWP22,piranha-code} }
      \addplot[smooth,mark=*,blue] plot coordinates {
        (1,1072.90) (1000,1002646.99 ) (10000,7268181.36) (100000,17547892.59) (1000000,17062429.72 )
      };
      \addlegendentry{Bicoptor 2.0 DReLU ($\ell_x=7$)}
      \end{axis}
      \end{tikzpicture}
      \begin{tikzpicture}
        \footnotesize
        \begin{axis}[
          height=3cm, width= 4.4cm, axis lines=middle,
          axis line style={->}, 
          ymin=0, ymax=6000000,
          xmode=log,
          x label style={at={(axis description cs:0.5,-0.25)},anchor=north},
          y label style={at={(axis description cs:-0.2,0.5)},rotate=90,anchor=south},
          title={LAN2},
          title style={at={(axis description cs:0.5,0.9)},anchor=north},
            xlabel=Batch size,
            ylabel=Throughput rate (ops),
            legend style={at={(1.5,1.2)},anchor=east}, legend columns=3]
        
        \addplot[smooth,mark=-,gray] plot coordinates {
            (1,111.23 ) (1000, 93258.35 ) (10000,492135.67 ) (100000,672336.71 ) (1000000,707018.57 )
        };
        \addplot[smooth,mark=*,blue] plot coordinates {
          (1,743.50 ) (1000,713210.80 ) (10000,3898498.69 ) (100000,5863383.17 ) (1000000,4648006.47)
        };
        \end{axis}
        \end{tikzpicture}
        \begin{tikzpicture}
          \footnotesize
          \begin{axis}[
            height=3cm, width= 4.4cm, axis lines=middle,
            axis line style={->}, 
            ymin=0, ymax=600000,
            xmode=log,
            x label style={at={(axis description cs:0.5,-0.25)},anchor=north},
            y label style={at={(axis description cs:-0.2,0.5)},rotate=90,anchor=south},
            title={WAN},
            title style={at={(axis description cs:0.5,0.9)},anchor=north},
              xlabel=Batch size,
              ylabel=Throughput rate (ops),
              legend style={at={(1.5,1.2)},anchor=east}, legend columns=3]
          
          \addplot[smooth,mark=-,gray] plot coordinates {
              (1,3.50) (1000,3393.71 ) (10000,25330.24) (100000,58377.80) (1000000,80134.63)
          };
          \addplot[smooth,mark=*,blue] plot coordinates {
            (1,24.40) (1000,23871.19 ) (10000,184700.86) (100000,430429.65) (1000000,526221.62)
          };
          \end{axis}
          \end{tikzpicture}
      
      \begin{tikzpicture}
      \footnotesize
      \begin{axis}[
        height=3cm, width= 4.4cm, axis lines=middle,
        axis line style={->}, 
        ymin=0, ymax=15000000,
        xmode=log,
        x label style={at={(axis description cs:0.5,-0.25)},anchor=north},
        y label style={at={(axis description cs:-0.2,0.5)},rotate=90,anchor=south},
        title={LAN1},
        title style={at={(axis description cs:0.5,0.9)},anchor=north},
          xlabel=Batch size,
          ylabel=Throughput rate (ops),
          legend style={at={(-0.5,0.5)},anchor=east}, legend columns=1]
      \addplot[smooth,mark=-,gray] plot coordinates {
          (1,137.64 ) (1000,128313.54 ) (10000,774263.48 ) (100000,1525173.76 ) (1000000,1666266.76 )
      };
      \addlegendentry{P-Falcon ReLU~\cite{WWP22,piranha-code} }
      \addplot[smooth,mark=*,blue] plot coordinates {
        (1,658.14 ) (1000,621596.76 ) (10000,4395585.07 ) (100000,10041441.03 ) (1000000,11088207.80 )
      };
      \addlegendentry{Bicoptor 2.0 ReLU ($\ell_x=7$)}
      \end{axis}
      \end{tikzpicture}
      \begin{tikzpicture}
        \footnotesize
        \begin{axis}[
          height=3cm, width= 4.4cm, axis lines=middle,
          axis line style={->}, 
          ymin=0, ymax=4000000,
          xmode=log,
          x label style={at={(axis description cs:0.5,-0.25)},anchor=north},
          y label style={at={(axis description cs:-0.2,0.5)},rotate=90,anchor=south},
          title={LAN2},
          title style={at={(axis description cs:0.5,0.9)},anchor=north},
            xlabel=Batch size,
            ylabel=Throughput rate (ops),
            legend style={at={(1.5,1.2)},anchor=east}, legend columns=3]
        
        \addplot[smooth,mark=-,gray] plot coordinates {
            (1,99.75) (1000, 93541.87  ) (10000, 451905.91 ) (100000, 656021.62 ) (1000000, 653163.60)
        };
        \addplot[smooth,mark=*,blue] plot coordinates {
          (1,444.10) (1000, 426070.40) (10000, 2174901.21 ) (100000, 3197114.92 ) (1000000, 2946653.78 )
        };
        \end{axis}
        \end{tikzpicture}
        \begin{tikzpicture}
          \footnotesize
          \begin{axis}[
            height=3cm, width= 4.4cm, axis lines=middle,
            axis line style={->}, 
            ymin=0, ymax=400000,
            xmode=log,
            x label style={at={(axis description cs:0.5,-0.25)},anchor=north},
            y label style={at={(axis description cs:-0.2,0.5)},rotate=90,anchor=south},
            title={WAN},
            title style={at={(axis description cs:0.5,0.9)},anchor=north},
              xlabel=Batch size,
              ylabel=Throughput rate (ops),
              legend style={at={(1.5,1.2)},anchor=east}, legend columns=3]
          
          \addplot[smooth,mark=-,gray] plot coordinates {
              (1,3.48) (1000,3188.61) (10000,22102.62) (100000,46057.48) (1000000,72840.64)
          };
          \addplot[smooth,mark=*,blue] plot coordinates {
            (1,12.31) (1000,12060.67) (10000,114435.72) (100000,192806.77) (1000000,279506.84)
          };
          \end{axis}
          \end{tikzpicture}
      
  \caption{Performance comparisons of P-Falcon~\cite{WWP22,piranha-code} and Bicoptor 2.0 DReLU and ReLU protocols on the different networks and batch sizes. (Graph)}
  \label{fig:relu}
  \end{figure*}
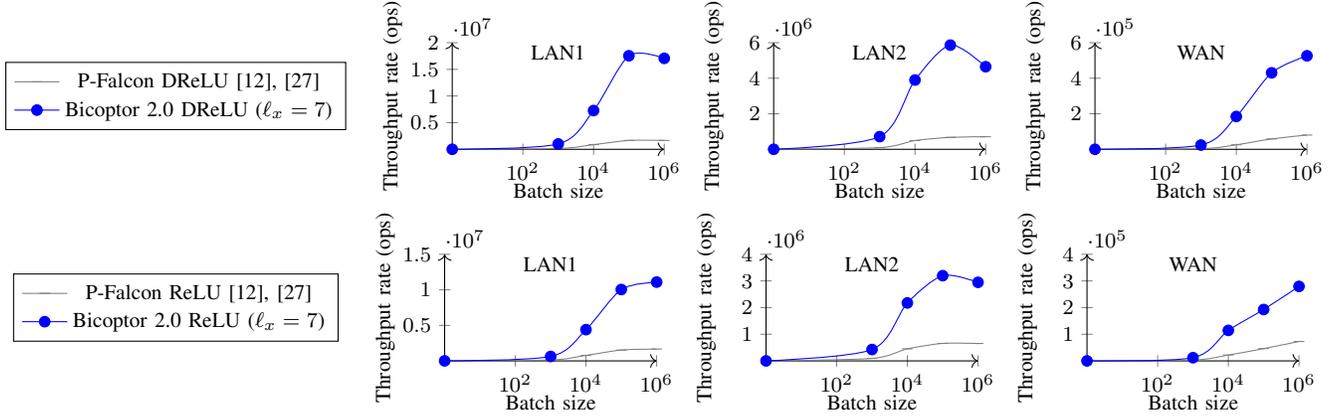
\begin{table*}[t!]
  \caption{Performance comparisons of P-Falcon~\cite{WWP22,piranha-code} and Bicoptor 2.0 DReLU protocols on the different networks and batch sizes. (ops) for operations per second.}
  \label{tab:drelu}
  \centering
  \linespread{1.25}
  \footnotesize
\begin{tabular*}{17.5cm}{@{\extracolsep{\fill}} p{0.5cm} l p{1cm} p{1.2cm} p{1cm} p{1.2cm} p{1cm} p{1.2cm}}
  \toprule
  \multirow{2}*{\textbf{Batch}} & \multirow{2}*{\textbf{Protocol}} & 
  \multicolumn{2}{c}{\textbf{LAN1}} & \multicolumn{2}{c}{\textbf{LAN2}} & \multicolumn{2}{c}{\textbf{WAN}}\\
  & & \textbf{Time} & \textbf{Thr. (ops)} & \textbf{Time} & \textbf{Thr. (ops)} & \textbf{Time} & \textbf{Thr. (ops)} \\
  \hline
  \multirow{2}*{1} 
  & P-Falcon DReLU~\cite{WWP22,piranha-code}
  & 6134.61$\mu$s & 163.01 & 8990.29$\mu$s & 111.23 & 285929$\mu$s & 3.50 \\
  & Bicoptor 2.0 DReLU ($\ell_x=7$) 
  & 932.05$\mu$s & 1072.90 & 1344.99$\mu$s & 743.50 & 40988.2$\mu$s & 24.40  \\
  \hline
  \multirow{2}*{$10^3$} 
  & P-Falcon DReLU~\cite{WWP22,piranha-code}
  & 6586.94$\mu$s & 151815.56 & 10722.9$\mu$s & 93258.35 & 294663$\mu$s & 3393.71 \\
  & Bicoptor 2.0 DReLU ($\ell_x=7$) 
  & 997.36$\mu$s & 1002646.99 & 1402.11$\mu$s & 713210.80 & 41891.5$\mu$s & 23871.19\\
  \hline
  \multirow{2}*{$10^4$} 
  & P-Falcon DReLU~\cite{WWP22,piranha-code}
  & 11429.2$\mu$s & 874951.88 & 20319.6$\mu$s & 492135.67& 394785$\mu$s & 25330.24 \\
  & Bicoptor 2.0 DReLU ($\ell_x=7$) 
  & 1375.86$\mu$s & 7268181.36 & 2565.09$\mu$s & 3898498.69 & 54141.6$\mu$s & 184700.86 \\
  \hline
  \multirow{2}*{$10^5$} 
  & P-Falcon DReLU~\cite{WWP22,piranha-code} 
  & 60192$\mu$s & 1661350.35 & 148735$\mu$s & 672336.71  & 1712980$\mu$s & 58377.80 \\
  & Bicoptor 2.0 DReLU ($\ell_x=7$) 
  & 5698.69$\mu$s & 17547892.59  & 17055$\mu$s & 5863383.17  & 232326$\mu$s & 430429.65 \\
  \hline
  \multirow{2}*{$10^6$} 
  & P-Falcon DReLU~\cite{WWP22,piranha-code} 
  & 600084$\mu$s & 1666433.37 & 1414390$\mu$s & 707018.57  & 12479000$\mu$s & 80134.63  \\
  & Bicoptor 2.0 DReLU ($\ell_x=7$) 
  & 58608.3$\mu$s & 17062429.72  & 215146$\mu$s & 4648006.47& 1900340$\mu$s & 526221.62 \\
  \bottomrule
  \end{tabular*}
\end{table*}
\begin{table*}[t!]
  \caption{Performance comparisons of P-Falcon~\cite{WWP22,piranha-code} and Bicoptor 2.0 ReLU protocols on the different networks and batch sizes. (ops) for operations per second.}
  \label{tab:relu}
  \centering
  \linespread{1.25}
  \footnotesize
\begin{tabular*}{17.5cm}{@{\extracolsep{\fill}} p{0.5cm} l p{1cm} p{1.2cm} p{1cm} p{1.2cm} p{1cm} p{1.2cm}}
  \toprule
  \multirow{2}*{\textbf{Batch}} & \multirow{2}*{\textbf{Protocol}} & 
  \multicolumn{2}{c}{\textbf{LAN1}} & \multicolumn{2}{c}{\textbf{LAN2}} & \multicolumn{2}{c}{\textbf{WAN}}\\
  & & \textbf{Time} & \textbf{Thr. (ops)} & \textbf{Time} & \textbf{Thr. (ops)} & \textbf{Time} & \textbf{Thr. (ops)} \\
  \hline
  \multirow{2}*{1} 
  & P-Falcon ReLU~\cite{WWP22,piranha-code} 
  & 7265.09$\mu$s & 137.64 & 10024.9$\mu$s & 99.75 & 287178$\mu$s & 3.48 \\
  & Bicoptor 2.0 ReLU ($\ell_x=7$) 
  & 1519.43$\mu$s & 658.14 & 2251.72$\mu$s & 444.10 & 81208.4$\mu$s & 12.31 \\
  \hline
  \multirow{2}*{$10^3$} 
  & P-Falcon ReLU~\cite{WWP22,piranha-code} 
  & 7793.41$\mu$s & 128313.54 & 10690.4$\mu$s & 93541.87 & 313616$\mu$s & 3188.61 \\
  & Bicoptor 2.0 ReLU ($\ell_x=7$) 
  & 1608.76$\mu$s & 621596.76 & 2347.03$\mu$s & 426070.40 & 82914.1$\mu$s & 12060.67 \\
  \hline
  \multirow{2}*{$10^4$} 
  & P-Falcon ReLU~\cite{WWP22,piranha-code} 
  & 12915.5$\mu$s & 774263.48 & 22128.5$\mu$s & 451905.91 & 452435$\mu$s & 22102.62 \\
  & Bicoptor 2.0 ReLU ($\ell_x=7$) 
  & 2275.01$\mu$s & 4395585.07 & 4597.91$\mu$s & 2174901.21 & 87385.3$\mu$s & 114435.72 \\
  \hline
  \multirow{2}*{$10^5$} 
  & P-Falcon ReLU~\cite{WWP22,piranha-code} 
  & 65566.3$\mu$s & 1525173.76 & 152434$\mu$s & 656021.62 & 2171200$\mu$s & 46057.48 \\
  & Bicoptor 2.0 ReLU ($\ell_x=7$) 
  & 9958.73$\mu$s & 10041441.03 & 31278.2$\mu$s & 3197114.92 & 518654$\mu$s & 192806.77 \\
  \hline
  \multirow{2}*{$10^6$} 
  & P-Falcon ReLU~\cite{WWP22,piranha-code} 
  & 600144$\mu$s & 1666266.76 & 1531010$\mu$s & 653163.60 & 13728600$\mu$s & 72840.64 \\
  & Bicoptor 2.0 ReLU ($\ell_x=7$) 
  & 90185.9$\mu$s & 11088207.80 & 339368$\mu$s & 2946653.78 & 3577730$\mu$s & 279506.84 \\
  \bottomrule
  \end{tabular*}
\end{table*}

\begin{table*}[t!]
  \caption{Performance comparisons of P-Falcon~\cite{WWP22,piranha-code} and Bicoptor 2.0 ReLU protocols on different models and batch sizes. We measure the total inference time ($\text{Time}_{\text{Total}}$) and the time taken by ReLU operation ($\text{Time}_{\text{ReLU}}$) for each protocol. $\text{Time}_{\text{comm}}$ is the total time cost by communication. We also compare the total inference time with plaintext inference time to evaluate the efficiency of our protocol. The batch size is selected to ensure each participant spends around 8 GiB GPU memory.}
  \label{tab:performance}
  \centering
  \linespread{1.25}
  \footnotesize
\begin{tabular*}{17.5cm}{@{\extracolsep{\fill}} p{0.9cm} l p{0.7cm} 
    p{0.7cm} p{0.7cm} p{0.7cm} p{0.7cm} p{0.7cm} p{0.7cm} p{0.7cm} p{0.7cm} p{0.7cm} p{1cm}}
  \toprule
  \multirow{2}*{\textbf{Model}} & \multirow{2}*{\textbf{Protocol}} & \textbf{Batch} &  \multicolumn{3}{c}{LAN1} & \multicolumn{3}{c}{LAN2} & \multicolumn{3}{c}{WAN} & Plaintext\\
  & & \textbf{size} 
  & $\text{Time}_{\text{ReLU}}$ & $\text{Time}_{\text{comm}}$ & $\text{Time}_{\text{Total}}$ 
  & $\text{Time}_{\text{ReLU}}$ & $\text{Time}_{\text{comm}}$ & $\text{Time}_{\text{Total}}$ 
  & $\text{Time}_{\text{ReLU}}$ & $\text{Time}_{\text{comm}}$ & $\text{Time}_{\text{Total}}$ 
  & $\text{Time}_{\text{Total}}$ \\
  \hline
  \multirow{3}{0.8cm}{CIFAR10\_\\AlexNet}
  & P-Falcon~\cite{WWP22,piranha-code} & \multirow{3}*{1650}
  & 13.72s & 13.42s & 16.72s 
  & 31.69s & 33.70s & 36.89s 
  & 262.38s & 294.13s & 297.45s & \multirow{3}*{0.32s}\\
  & Bicoptor 2.0 (RSS, $\ell_x=7$) &
  & 2.12s & 3.07s & 5.00s
  & 7.20s & 10.36s & 12.32s
  & 67.80s & 97.86s & 99.83s \\
  & Bicoptor 2.0 (UBL, $\ell_x=7$) &
  & 2.11s & 2.44s & 4.32s 
  & 7.81s & 9.17s & 11.07s  
  & 70.64s & 82.30s & 84.12s & \\
  \hline
  \multirow{3}{0.8cm}{Tiny\_\\AlexNet}
  & P-Falcon~\cite{WWP22,piranha-code} & \multirow{3}*{510}
  & 24.47s & 24.04s & 30.47s 
  & 54.77s & 58.26s & 64.71s 
  & 450.97s & 506.83s & 513.48s & \multirow{3}*{0.10s}\\
  & Bicoptor 2.0 (RSS, $\ell_x=7$) &
  & 3.82s & 5.71s & 10.02s
  & 12.37s & 17.90s & 22.21s 
  & 120.92s & 175.15s & 179.53s \\
  & Bicoptor 2.0 (UBL, $\ell_x=7$) &
  & 3.76s & 4.41s & 8.60s 
  & 13.17s & 15.54s & 19.76s 
  & 123.82s & 148.13s & 152.15s & \\
  \hline
  \multirow{3}{0.8cm}{CIFAR10\_\\VGG16}
  & P-Falcon~\cite{WWP22,piranha-code} & \multirow{3}*{240}
  & 46.11s & 45.08s & 54.28s 
  & 106.55s & 112.90s & 122.01s 
  & 861.32s & 959.08s & 968.43s  & \multirow{3}*{0.55s}\\
  & Bicoptor 2.0 (RSS, $\ell_x=7$) &
  & 6.90s & 10.07s & 15.17s 
  & 23.53s & 33.82s & 38.92s 
  & 229.07s & 331.42s & 336.64s \\
  & Bicoptor 2.0 (UBL, $\ell_x=7$) &
  & 7.08s & 8.41s & 13.22s 
  & 25.89s & 15.39s & 30.89s 
  & 239.87s & 287.89s & 292.40s & \\
  \hline
  \multirow{3}{0.8cm}{Tiny\_\\VGG16}
  & P-Falcon~\cite{WWP22,piranha-code} & \multirow{3}*{60}
  & 46.78s & 45.88s & 55.02s 
  & 106.27s & 112.51s & 121.62s 
  & 859.60s & 958.44s & 967.74s & \multirow{3}*{0.15s}\\
  & Bicoptor 2.0 (RSS, $\ell_x=7$) &
  & 6.90s & 10.24s & 15.35s
  & 23.69s & 34.00s & 39.12s 
  & 229.53s & 330.88s & 336.09s \\
  & Bicoptor 2.0 (UBL, $\ell_x=7$) &
  & 7.04s & 8.19s & 13.01s 
  & 25.00s & 29.37s & 34.208s 
  & 239.06s & 286.53s & 291.05s  & \\
  \bottomrule
  \end{tabular*}
\end{table*}

\subsection{E2E PPML Inference Experiments} \label{sub:e2etest} 
\begin{figure}[t!]
  \centering
  \begin{tikzpicture}
    \footnotesize
    \begin{axis}[
      xbar stacked,
      legend pos = south east,
      legend style={font = \tiny},
      legend columns = 2,
      ytick={0,1,2},  
      yticklabels={
        Bicoptor 2.0 (UBL) (48.84\%),
        Bicoptor 2.0 (RSS) (42.40\%),
        P-Falcon~\cite{WWP22,piranha-code} (82.06\%),
      },
      y tick label style={font = \footnotesize},
      xmin=0,
      height=2.5cm, 
      width=5cm, 
      bar width=0.2cm, 
      x tick label style={font = \footnotesize},
      x label style={at={(axis description cs:0.1, -0.15)},anchor=south, font=\footnotesize},
      title={(a) CIFAR10\_AlexNet\_1650 },
      title style={at={(axis description cs:0.3,1.2)},anchor=north},
      legend style={at={(1.05,1.6)},anchor=east}]
      \addplot coordinates
      { (13.72,2) (2.12,1) (2.11,0) };
      \addplot coordinates
      { (16.72-13.72,2) (5.00-2.12,1) (4.32-2.11,0) };
      \legend{ReLU latency (s), Other latency (s)}
      \end{axis}
  \end{tikzpicture}
  \begin{tikzpicture}
    \footnotesize
    \begin{axis}[
      xbar stacked,
      legend pos = south east,
      legend style={font = \tiny},
      legend columns = 2,
      ytick={0,1,2},  
      yticklabels={
        Bicoptor 2.0 (UBL) (43.72\%),
        Bicoptor 2.0 (RSS) (38.12\%),
        P-Falcon~\cite{WWP22,piranha-code} (80.31\%),
      },
      y tick label style={font = \footnotesize},
      xmin=0,
      height=2.5cm, 
      width=5cm, 
      bar width=0.2cm, 
      x tick label style={font = \footnotesize},
      x label style={at={(axis description cs:0.1, -0.15)},anchor=south, font=\footnotesize},
      title={(b) Tiny\_AlexNet\_510 },
      title style={at={(axis description cs:0.3,1.2)},anchor=north},
      legend style={at={(1.05,1.4)},anchor=east}]
      \addplot coordinates
      { (24.47,2) (3.82,1) (3.76,0)};
      \addplot coordinates
      { (30.47-24.47,2) (10.02-3.82,1) (8.60-3.76,0) };
      \end{axis}
  \end{tikzpicture}
  \begin{tikzpicture}
    \footnotesize
    \begin{axis}[
      xbar stacked,
      legend pos = south east,
      legend style={font = \tiny},
      legend columns = 2,
      ytick={0,1,2},  
      yticklabels={
        Bicoptor 2.0 (UBL) (53.56\%),
        Bicoptor 2.0 (RSS) (45.48\%),
        P-Falcon~\cite{WWP22,piranha-code} (84.95\%),
      },
      y tick label style={font = \footnotesize},
      xmin=0,
      height=2.5cm, 
      width=5cm, 
      bar width=0.2cm, 
      x tick label style={font = \footnotesize},
      x label style={at={(axis description cs:0.1, -0.15)},anchor=south, font=\footnotesize},
      title={(c) CIFAR10\_VGG16\_240 },
      title style={at={(axis description cs:0.3,1.2)},anchor=north},
      legend style={at={(1.05,1.4)},anchor=east}]
      \addplot coordinates
      { (46.11,2) (6.90,1) (7.08,0) };
      \addplot coordinates
      { (54.28-46.11,2) (15.17-6.90,1) (13.22-7.08,0) };
      \end{axis}
  \end{tikzpicture}
  \begin{tikzpicture}
    \footnotesize
    \begin{axis}[
      xbar stacked,
      legend pos = south east,
      legend style={font = \tiny},
      legend columns = 2,
      ytick={0,1,2},  
      yticklabels={
        Bicoptor 2.0 (UBL) (54.11\%),
        Bicoptor 2.0 (RSS) (44.95\%),
        P-Falcon~\cite{WWP22,piranha-code} (85.02\%),
      },
      y tick label style={font = \footnotesize},
      xmin=0,
      height=2.5cm, 
      width=5cm, 
      bar width=0.2cm, 
      x tick label style={font = \footnotesize},
      x label style={at={(axis description cs:0.1, -0.15)},anchor=south, font=\footnotesize},
      title={(d) Tiny\_VGG16\_60 },
      title style={at={(axis description cs:0.3,1.2)},anchor=north},
      legend style={at={(1.05,1.4)},anchor=east}]
      \addplot coordinates
      { (46.78,2) (6.90,1) (7.04,0) };
      \addplot coordinates
      { (55.02-46.78,2) (15.35-6.90,1) (13.01-7.04,0) };
      \end{axis}
  \end{tikzpicture}
  \caption{Runtime comparison of P-Falcon~\cite{WWP22,piranha-code} and Bicoptor 2.0 ReLU protocols and the inference for various models in LAN1. In Bicoptor 2.0, $\ell_x=7$ for DReLU, and both UBL and RSS modes are evaluated. The batch size is selected to ensure each participant spends around 8 GiB GPU memory.}
  \label{fig:e2eperformance}
\end{figure}
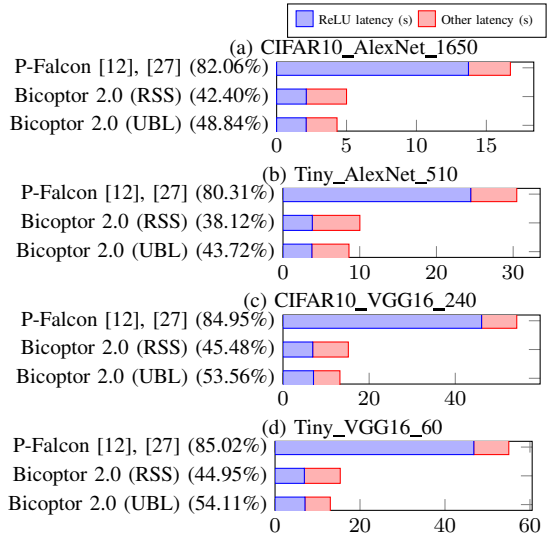

We conduct experiments on E2E PPML inference using our fully optimized ReLU protocol, comparing it to the original P-Falcon~\cite{WWP22,piranha-code} on four different models under different network environments. We use UBL instead of RSS in the linear layers because it is slightly faster than using RSS. The results of the experiments are presented in Fig.~\ref{fig:e2eperformance} and Tab.~\ref{tab:performance}. We achieve a 3-4x improvement under different network conditions for different models. For example, by using our ReLU protocol in CIFAR10\_AlexNet under LAN1, we achieved a total inference time of $4.4s$ while the original Piranha~\cite{WWP22,piranha-code} requires $16.72s$, demonstrating a 4x of improvement. This improvement narrows the performance gap between the PPML inference and plaintext inference to only a factor of 20 in model CIFAR10\_AlexNet, and between 20-100x in other models (Tab.~\ref{tab:performance2}).

To further demonstrate the performance improvement brought about by our optimized ReLU protocol from a different perspective, we would like to recall that the cost of ReLU layer accounts for around $80\%$ of the total inference cost. With our optimizations (on both UBL and RSS), this portion has been reduced to around $50\%$. Fig.~\ref{fig:e2eperformance} illustrates the percentage of ReLU cost for different models under LAN1 when using our optimized ReLU protocol.

\begin{table}[t!]
  \caption{Performance comparison of E2E PPML inference and PyTorch plaintext inference in LAN1 with batch size 128.}
  \label{tab:performance2}
  \centering
  \linespread{1.25}
  \footnotesize
\begin{tabular*}{8.5cm}{@{\extracolsep{\fill}} p{0.8cm}  l p{1.7cm} p{1cm} }
  \toprule
  \textbf{Model} & \textbf{Protocol} &  $\text{Time}_{\text{Total}}^{\text{PPML}}$ & $\text{Time}_{\text{Total}}^{\text{PlainML}}$ \\
  \hline
  CIFAR10\_ & P-Falcon~\cite{WWP22,piranha-code} & 1.75s & \multirow{2}*{0.03s} \\
  AlexNet  & Bicoptor 2.0 (UBL, $\ell_x=7$) & 0.61s & \\
  \hline
  Tiny\_  & P-Falcon~\cite{WWP22,piranha-code} & 7.81s & \multirow{2}*{0.03s} \\
  AlexNet & Bicoptor 2.0 (UBL, $\ell_x=7$)  & 2.41s & \\
  \hline
  CIFAR10\_ & P-Falcon~\cite{WWP22,piranha-code} & 30.48s & \multirow{2}*{0.29s} \\
  VGG16 & Bicoptor 2.0 (UBL, $\ell_x=7$)  & 7.12s & \\
  \hline
  Tiny\_ & P-Falcon~\cite{WWP22,piranha-code} & out of memory & \multirow{2}*{0.29s} \\
  VGG16 & Bicoptor 2.0 (UBL, $\ell_x=7$)  & 28.80s & \\
  \bottomrule
  \end{tabular*}
\end{table}

\section{Conclusion} \label{sec:conclusion}
Overall, after conducting extensive experiments and theoretical analysis, we provide two guidelines: one for selecting appropriate precision when using probabilistic truncation to avoid accuracy degradation in inference, and another for how to perform DReLU computations using partial precision, i.e., key bits, to improve DReLU performance under different models. The use of key bits is not only applicable to Bicoptor~\cite{ZWC+-23}, but also to other works. 

In addition to the guidelines, our work also proposes two novel protocols: the non-interactive deterministic truncation protocol and the non-interactive MPC-based modulo-switch protocol both do not require preprocessing. By leveraging these protocols, the performance of our end-to-end PPML inference is only 20x slower than that of plaintext inference, representing a qualitative leap in the performance of PPML.

\bibliographystyle{IEEEtran}
\bibliography{IEEEabrv,myref}

\appendix

\subsection{Proof of Theorem~\ref{thm:cut}} \label{app:sub:proof:thm:cut}
We introduce a few interesting properties of the cut function mentioned in Sect.~\ref{sec:analysis} by a few lemmas and corollaries, which would be summarised into Theorem~\ref{thm:cut}.

Lemma~\ref{lmm:cut} demonstrates how to transform the cut of the sum of two numbers into the cuts of the two numbers separately, and then sums them up when there is no modulo overflow.
\begin{lemma} \label{lmm:cut}
  For $\alpha,\beta \in \mathbb{Z}_{2^\ell}$, $\alpha+\beta\ \text{mod}\ 2^\ell \geq \alpha$, $\mathsf{bit} := \{0,1\}$, 
  \begin{itemize} [leftmargin=*]
    \item If $\alpha + \beta\ \text{mod}\ {2^\ell} \geq \alpha$,
    $\mathsf{cut}(\alpha+\beta\ \text{mod}\ 2^\ell,k) = \mathsf{cut}(\alpha,k) + \mathsf{cut}(\beta,k) + \mathsf{bit}\ \text{mod}\ 2^\ell$.
    \item If $\alpha - \beta\ \text{mod}\ {2^\ell} \leq \alpha$,
    $\mathsf{cut}(\alpha-\beta\ \text{mod}\ 2^\ell,k) = \mathsf{cut}(\alpha,k) - \mathsf{cut}(\beta,k) - \mathsf{bit}\ \text{mod}\ 2^\ell$.
  \end{itemize}
\end{lemma}

Lemma~\ref{lmm:cut} is a special case of Theorem~\ref{thm:cut}, please refer to Case (a) in the proof of Theorem~\ref{thm:cut}.

With Lemma~\ref{lmm:cut} when considering $\alpha:=2^\ell$ and $\beta:=\gamma$, we have the following useful property:
\begin{corollary} \label{clr:cut}
  $\mathsf{cut}(- \gamma\ \text{mod}\ 2^\ell,k) = \mathsf{cut}(2^\ell,k) - \mathsf{cut}(\gamma,k) - 1\ \text{mod}\ 2^\ell.$
\end{corollary}

Moreover, Lemma~\ref{lmm:cut2} describes the property of cut of the sum of two numbers when there is modulo overflow.
\begin{lemma} \label{lmm:cut2}
If $\alpha+\beta\ \text{mod}\ 2^\ell < \alpha$, $\mathsf{bit} := \{0,1\}$, 
\begin{itemize} [leftmargin=*]
  \item If $\alpha + \beta\ \text{mod}\ {2^\ell} < \alpha$,
  $\mathsf{cut}(\alpha+\beta\ \text{mod}\ 2^\ell,k) = \mathsf{cut}(\alpha,k) + \mathsf{cut}(\beta,k) + \mathsf{bit} - \mathsf{cut}(2^\ell,k)$.
  \item If $\alpha - \beta\ \text{mod}\ {2^\ell} > \alpha$,
  $\mathsf{cut}(\alpha-\beta\ \text{mod}\ 2^\ell,k) = \mathsf{cut}(\alpha,k) - \mathsf{cut}(\beta,k) - \mathsf{bit} - \mathsf{cut}(2^\ell,k)$.
\end{itemize}
\end{lemma}
\begin{IEEEproof}
  $\mathsf{cut}(\alpha+\beta\ \text{mod}\ 2^\ell,k) = \mathsf{cut}(\alpha + \beta - 2^\ell,k) 
= \mathsf{cut}(\alpha - (2^\ell - \beta),k) = \mathsf{cut}(\alpha,k) - \mathsf{cut}(2^\ell - \beta,k) - \mathsf{bit} = \mathsf{cut}(\alpha,k) + \mathsf{cut}(\beta,k) + 1 - \mathsf{cut}(2^\ell,k) - \mathsf{bit} = \mathsf{cut}(\alpha,k) + \mathsf{cut}(\beta,k) + \mathsf{bit} - \mathsf{cut}(2^\ell,k)$.

  By Lemma~\ref{lmm:cut} we have
  \begin{align*}
    &\mathsf{cut}(\alpha,k) - (\mathsf{cut}(2^\ell,k) - \mathsf{cut}(\beta,k) - 1) - \mathsf{bit}\\
  = &\mathsf{cut}(\alpha,k) - \mathsf{cut}(2^\ell,k) + \mathsf{cut}(\beta,k) + 1 - \mathsf{bit}\\
  = &\mathsf{cut}(\alpha,k) + \mathsf{cut}(\beta,k) - \mathsf{cut}(2^\ell,k) + \mathsf{bit};
  \end{align*}
  If $\alpha-\beta > \alpha$, then
  $\mathsf{cut}(\alpha-\beta\ \text{mod}\ 2^\ell,k) = \mathsf{cut}(\alpha-\beta + 2^\ell,k)
  = \mathsf{cut}(\alpha + ({2^\ell} - \beta),k)$. By Corollary~\ref{clr:cut} we have
  \begin{align*}
   &\mathsf{cut}(\alpha,k) + (\mathsf{cut}(2^\ell,k) - \mathsf{cut}(\beta,k) - 1) + \mathsf{bit}\\
  =&\mathsf{cut}(\alpha,k) + \mathsf{cut}(2^\ell,k) - \mathsf{cut}(\beta,k) - 1 + \mathsf{bit}\\
  =&\mathsf{cut}(\alpha,k) - \mathsf{cut}(\beta,k) + \mathsf{cut}(2^\ell,k) - \mathsf{bit};
  \end{align*}
\end{IEEEproof}

We now summarise Lemma~\ref{lmm:cut2} and Lemma~\ref{lmm:cut} into Theorem~\ref{thm:cut}. The proof of Theorem~\ref{thm:cut} is given as follows.
\begin{IEEEproof}
  Considering $\alpha:=\alpha^\prime\cdot 2^k + \alpha^{\prime\prime}$ and $\beta:=\beta^\prime\cdot 2^k + \beta^{\prime\prime}$, where $\alpha^\prime,\beta^\prime\in\mathbb{Z}_{2^{\ell-k}}$ and $\alpha^{\prime\prime},\beta^{\prime\prime}\in\mathbb{Z}_{2^k}$.

  Case (a): For $\alpha+\beta=(\alpha^\prime+\beta^\prime)\cdot 2^k + \alpha^{\prime\prime}+\beta^{\prime\prime} < 2^\ell$. 
  \begin{itemize} [leftmargin=*]
    \item If $\alpha^{\prime\prime}+\beta^{\prime\prime}<2^k$, then $\mathsf{cut}(\alpha+\beta,k)=\alpha^\prime+\beta^\prime=\mathsf{cut}(\alpha,k)+\mathsf{cut}(\beta,k)$. 
    \item If $\alpha^{\prime\prime}+\beta{\prime\prime} \geq 2^k$, then $\alpha+\beta=(\alpha^\prime+\beta^\prime + 1)\cdot 2^k + (\alpha^{\prime\prime}+\beta^{\prime\prime}-2^k)$. Hence, $\mathsf{cut}(\alpha+\beta,k)=\alpha^\prime+\beta^\prime + 1=\mathsf{cut}(\alpha,k)+\mathsf{cut}(\beta,k)+1$. 
  \end{itemize}

  Case (b): For $\alpha-\beta=(\alpha^\prime-\beta^\prime)\cdot 2^k + \alpha^{\prime\prime}-\beta^{\prime\prime} \geq 0$. 
  \begin{itemize} [leftmargin=*]
    \item If $\alpha^{\prime\prime}-b^{\prime\prime}\geq 0$, then $\mathsf{cut}(\alpha-\beta,k)=\alpha^\prime-\beta^\prime=\mathsf{cut}(\alpha,k)-\mathsf{cut}(\beta,k)$. 
    \item If $\alpha^{\prime\prime}-b^{\prime\prime}<0$, then $\alpha-\beta=(\alpha^\prime-\beta^\prime - 1)\cdot 2^k + (\alpha^{\prime\prime}-\beta^{\prime\prime}+2^k)$. Hence, $\mathsf{cut}(\alpha-\beta,k)=\alpha^\prime-\beta^\prime - 1=\mathsf{cut}(\alpha,k)-\mathsf{cut}(\beta,k)-1$.
  \end{itemize}

  Case (c): For $\alpha+\beta=(\alpha^\prime+\beta^\prime)\cdot 2^k + \alpha^{\prime\prime}+\beta^{\prime\prime} \geq 2^\ell$.\\
  Since $\alpha+\beta\ \text{mod}\ 2^\ell<\alpha$ and $\alpha,\beta\in\mathbb{Z}_{2^\ell}$, 
  we have:
  \begin{align*}  
    &\mathsf{cut}(\alpha+\beta\ \text{mod}\ 2^\ell,k)=\mathsf{cut}(\alpha+\beta - 2^\ell,k)
  \end{align*}
  Since $\alpha+\beta - 2^\ell = (\alpha^{\prime}+\beta^{\prime}-2^{\ell-k})\cdot 2^k + \alpha^{\prime\prime} + \beta^{\prime\prime}$, we have:
  \begin{align*}  
    &\mathsf{cut}(\alpha+\beta - 2^\ell,k) = \mathsf{cut}(\alpha,k) + \mathsf{cut}(\beta,k) - \mathsf{cut}(\ell,k)
  \end{align*}
  
  Case (d): For $\alpha-\beta=(\alpha^\prime+\beta^\prime)\cdot 2^k + \alpha^{\prime\prime}+\beta^{\prime\prime} < 0$.\\
  Since $\alpha-\beta > \alpha$ and $\alpha,\beta\in\mathbb{Z}_{2^\ell}$, 
  we have:
  \begin{align*}  
    &\mathsf{cut}(\alpha-\beta\ \text{mod}\ 2^\ell,k)=\mathsf{cut}(\alpha-\beta + 2^\ell,k)
  \end{align*}
  Since $\alpha-\beta + 2^\ell = (\alpha^{\prime}-\beta^{\prime}+2^{\ell-k})\cdot 2^k - \alpha^{\prime\prime} - \beta^{\prime\prime}$, we have:
  \begin{align*}  
    &\mathsf{cut}(\alpha-\beta + 2^\ell,k) = \mathsf{cut}(\alpha,k) - \mathsf{cut}(\beta,k) + \mathsf{cut}(\ell,k)
  \end{align*}
  

  All cases finalize the proof.
\end{IEEEproof}

\subsection{Proof of Theorem~\ref{thm:newcut2}} \label{app:sub:proof:thm:newcut2}
\begin{IEEEproof}
  Considering 
  $\alpha:=\alpha^\prime\cdot 2^{\ell-k_2} + \alpha^{\prime\prime}\cdot 2^{k_1} + \alpha^{\prime\prime\prime}$ and 
  $\beta:=\beta^\prime\cdot 2^{\ell-k_2} + \beta^{\prime\prime}\cdot 2^{k_1} + \beta^{\prime\prime\prime}$, 
  where 
  $\alpha^\prime,\beta^\prime\in\mathbb{Z}_{2^{k_1}}$,
  $\alpha^{\prime\prime},\beta^{\prime\prime}\in\mathbb{Z}_{2^{\ell-k_1-k_2}}$,
  and 
  $\alpha^{\prime\prime\prime},\beta^{\prime\prime\prime}\in\mathbb{Z}_{2^{k_2}}$.

  For $\alpha+\beta$, we have 
  $\mathsf{cut}(\alpha+\beta,k_1)\ \text{mod}\ 2^{k_1}
  =\mathsf{cut}(\alpha,k_1)+\mathsf{cut}(\beta,k_1)+\mathsf{bit}\ \text{mod}\ 2^{\ell-k_1}
  =(\alpha^\prime+\beta^\prime)\cdot 2^{\ell-k_1-k_2}+(\alpha^{\prime\prime}+\beta^{\prime\prime}+\mathsf{bit})\ \text{mod}\ 2^{\ell-k_1}$ from the Theorem~\ref{thm:newtrc2}. Hence, $\mathsf{cut}(\alpha+\beta,k_1,k_2)\ \text{mod}\ 2^{\ell-k_1-k_2}= \alpha^{\prime\prime}+\beta^{\prime\prime}+\mathsf{bit}\ \text{mod}\ 2^{\ell-k_1-k_2}$.
  
  For $\alpha-\beta$, we have 
  $\mathsf{cut}(\alpha-\beta,k_1)\ \text{mod}\ 2^{k_1}
  =\mathsf{cut}(\alpha,k_1)-\mathsf{cut}(\beta,k_1)-\mathsf{bit}\ \text{mod}\ 2^{\ell-k_1}
  =(\alpha^\prime-\beta^\prime)\cdot 2^{\ell-k_1-k_2}+(\alpha^{\prime\prime}-\beta^{\prime\prime}-\mathsf{bit})\ \text{mod}\ 2^{\ell-k_1}$ from the Theorem~\ref{thm:newtrc2}. Hence, $\mathsf{cut}(\alpha-\beta,k_1,k_2)\ \text{mod}\ 2^{\ell-k_1-k_2}= \alpha^{\prime\prime}-\beta^{\prime\prime}+\mathsf{bit}\ \text{mod}\ 2^{\ell-k_1-k_2}$.
\end{IEEEproof}

Similarly, Theorem~\ref{thm:newcut2} implies Proposition~\ref{pro:newcut2}.
\begin{proposition} \label{pro:newcut2}
  In a ring $\mathbb{Z}_{2^\ell}$, let $x\in[0,2^{\ell_x})\bigcup(2^\ell-2^{\ell_x},2^\ell)$, where $\ell>\ell_x + 1$, $\mathsf{bit}:=\{0,1\}$. We have the following properties:
  \begin{itemize} [leftmargin=*]
    \item For a positive $x$, $\xi:=x \in [0,2^{\ell_x}-1]$, then 
    \begin{align*}  
     &\mathsf{cut}(R+\xi,k_1,k_2)\ \text{mod}\ 2^{\ell-k_1-k_2}\\
    =&\mathsf{cut}(R,k)+\mathsf{cut}(\xi,k)+\mathsf{bit}\ \text{mod}\ 2^{\ell-k}.
    \end{align*}  
    \item For a negative $x$, $\xi:=2^\ell-x \ \text{mod}\ 2^\ell$, then 
    \begin{align*}  
      &\mathsf{cut}(R-\xi,k_1,k_2)\ \text{mod}\ 2^{\ell-k_1-k_2}\\
      =&\mathsf{cut}(R,k)-\mathsf{cut}(\xi,k)-\mathsf{bit}\ \text{mod}\ 2^{\ell-k}.
    \end{align*}  
  \end{itemize}
\end{proposition}

Finally, Proposition~\ref{pro:newcut2} implies Theorem~\ref{thm:newtrc2}, which concludes the correctness of Alg.~\ref{alg:newtrc2}.
\begin{theorem} \label{thm:newtrc2}
  In a ring $\mathbb{Z}_{2^\ell}$, let $x\in[0,2^{\ell_x})\bigcup(2^\ell-2^{\ell_x},2^\ell)$, where $\ell>\ell_x + 1$, $\mathsf{bit}:=\{0,1\}$. We have the following properties:
  \begin{itemize}[leftmargin=*]
      \item For a positive $x$, $\overline{\mathsf{trc}}(x,k_1,k_2)=\mathsf{cut}(\xi,k_1,k_2) + \mathsf{bit}\ \textsf{mod}\ 2^{\ell-k_1-k_2}$.
      \item For a negative $x$, $\overline{\mathsf{trc}}(x,k_1,k_2)=2^\ell-\mathsf{cut}(\xi,k_1,k_2) - \mathsf{bit}\ \textsf{mod}\ 2^{\ell-k_1-k_2}$.
  \end{itemize}
\end{theorem}

\begin{IEEEproof}
  If $x$ is positive, with Alg.~\ref{alg:newtrc2}, $[x]_0:=\xi + R$ and $[x]_1:=-R$, $P_0$ does $\mathsf{cut}([x]_0,k_1,k_2)=\mathsf{cut}(x+R\ \text{mod}\ 2^{\ell},k_1,k_2)$
  \begin{align*}
    =&\mathsf{cut}(x,k_1,k_2)+\mathsf{cut}(R,k_1,k_2)+\mathsf{bit}\\
    &-\mathsf{LT}(x + R\ \text{mod}\ 2^{\ell}, x)\cdot\mathsf{cut}(2^\ell,k_1,k_2)\ \text{mod}\ 2^{\ell-k_1-k_2}
  \end{align*}
  Note that $\mathsf{cut}(2^\ell,k_1,k_2) = 2^{\ell-k_1-k_2}\ \text{mod}\ 2^{\ell-k_1-k_2}=0\ \text{mod}\ 2^{\ell-k_1-k_2}$
  Then, $\mathsf{cut}([x]_0,k_1,k_2)$
  	$$=\mathsf{cut}(x,k_1,k_2)+\mathsf{cut}(R,k_1,k_2)+\mathsf{bit}\ \text{mod}\ 2^{\ell-k_1-k_2}.$$
  $P_1$ does $-\mathsf{cut}(R,k_1,k_2)\ \text{mod}\ 2^{\ell-k_1-k_2}$. Hence, 
  \begin{align*}
     &[\overline{\mathsf{trc}}(\xi,k_1,k_2)]_0+[\overline{\mathsf{trc}}(\xi,k_1,k_2)]_1\\
    =&\mathsf{cut}(x,k_1,k_2)+\mathsf{cut}(R,k_1,k_2)+\mathsf{bit} \\
     &-\mathsf{cut}(R,k_1,k_2)\ \text{mod}\ 2^{\ell-k_1-k_2}\\
    =&\mathsf{cut}(x,k_1,k_2)+\mathsf{bit}\ \text{mod}\ 2^{\ell-k_1-k_2}.
  \end{align*} 

  If $x$ is negative, then $[x]_0:=R-\xi$ and $[x]_1:=-R$. $P_0$ does 
  \begin{align*}
     &\mathsf{cut}(R-\xi,k_1,k_2)\\
    =&\mathsf{cut}(R,k_1,k_2)-\mathsf{cut}(\xi,k_1,k_2)-\mathsf{bit}\\
    &+\mathsf{LT}(R,R-\xi\ \text{mod}\ 2^{\ell})\cdot\mathsf{cut}(2^\ell,k_1,k_2)\ \text{mod}\ 2^{\ell-k_1-k_2},
  \end{align*} 
  and $P_1$ does $-\mathsf{cut}(R,k_1,k_2)\ \text{mod}\ 2^{\ell-k_1-k_2}$. Hence, 
  \begin{align*}
    &[\overline{\mathsf{trc}}(\xi,k_1,k_2)]_0+[\overline{\mathsf{trc}}(\xi,k_1,k_2)]_1\\
  =&-\mathsf{cut}(\xi,k_1,k_2)+\mathsf{bit}\ \text{mod}\ 2^{\ell-k_1-k_2}.
  \end{align*} 

\end{IEEEproof}

\subsection{Proof of Theorem~\ref{thm:pattern0}} \label{app:sub:proof:thm:pattern0}
We define $\lambda$ as the effective bit length, hence, $\xi_{\lambda-1}$ = 1. For $\lambda + 1 < \ell$. We also define the following notations $R_\lambda^{\prime\prime},R_\lambda^{\prime},R_{\lambda-1}^{\prime\prime},R_{\lambda-1}^{\prime},$\\$\xi_\lambda^{\prime\prime},\xi_\lambda^{\prime},\xi_{\lambda-1}^{\prime\prime},\xi_{\lambda-1}^{\prime}$ for the rest of Lemma
\begin{align*}
  &R = R_{\lambda}^{\prime}\cdot2^{\lambda} + R_\lambda^{\prime\prime},\ R = R_{\lambda-1}^{\prime}\cdot2^{\lambda-1} + R_{\lambda-1}^{\prime\prime},\\
  &\xi = \xi_{\lambda}^{\prime}\cdot2^{\lambda} + \xi_\lambda^{\prime\prime},\ \xi = \xi_{\lambda-1}^{\prime}\cdot2^{\lambda-1} + \xi_{\lambda-1}^{\prime\prime},
\end{align*} 
In addition,
$$R_\lambda^{\prime\prime} = R_{\lambda-1}^{\prime}\cdot2^{\lambda-1} + R_{\lambda-1}^{\prime\prime},\ \xi_\lambda^{\prime\prime} = \xi_{\lambda-1}^{\prime}\cdot2^{\lambda-1} + \xi_{\lambda-1}^{\prime\prime}.$$

Since $\xi_{\lambda-1}$ is the first effective bit which is always 1 and all bits to the left of $\xi_{\lambda-1}$ are 0, it is meaningless to consider cutting the bits to the left of $\xi_{\lambda-1}$. Hence, we only use $\overline{\mathsf{trc}}(\xi,\lambda-1)$ and $\overline{\mathsf{trc}}(\xi,\lambda)$ instead of $\overline{\mathsf{trc}}(\xi,\lambda-1, \ell - \lambda - k_2)$ and $\overline{\mathsf{trc}}(\xi,\lambda, \ell - \lambda - k_2 -1)$ for the following lemmas.

\begin{lemma} \label{lmm:pattern1}
  is a generalized version of Theorem~\ref{thm:smltrc}, and the proofs are similar.
  \begin{itemize} [leftmargin=*]
    \item If $x$ is positive, then $\overline{\mathsf{trc}}(\xi,k_1,k_2) = \mathsf{cut}(\xi,k_1,k_2) + \mathsf{bit}\ \text{mod}\ 2^{\ell-k_1-k_2}$. 
    \item If $x$ is negative, then $\overline{\mathsf{trc}}(\xi,k_1,k_2) = -\mathsf{cut}(\xi,k_1,k_2) - \mathsf{bit}\ \text{mod}\ 2^{\ell-k_1-k_2}$
  \end{itemize}
\end{lemma}

\begin{lemma} \label{lmm:pattern2}
  describes when exactly does $e_0$ occurs in Lemma~\ref{lmm:pattern1}. Considering $[x]_0:=\xi + R$ and $[x]_1:=-R$.
  \begin{itemize} [leftmargin=*]
  \item If $x$ is positive, then 
    \begin{enumerate} [leftmargin=*]
      \item $\overline{\mathsf{trc}}(\xi,k_1,k_2) = \mathsf{cut}(\xi,k_1,k_2)\ \text{mod}\ 2^{\ell-k_1-k_2}$ if $\xi^{\prime\prime} + R^{\prime\prime} < 2^\ell$. 
      \item $\overline{\mathsf{trc}}(\xi,k_1,k_2) = \mathsf{cut}(\xi,k_1,k_2) + 1\ \text{mod}\ 2^{\ell-k_1-k_2}$ if $\xi^{\prime\prime} + R^{\prime\prime} \geq 2^\ell$. 
    \end{enumerate}
  \item If $x$ is negative, then 
    \begin{enumerate} [leftmargin=*]
      \item $\overline{\mathsf{trc}}(\xi,k_1,k_2) = -\mathsf{cut}(\xi,k_1,k_2)\ \text{mod}\ 2^{\ell-k_1-k_2}$ if $R^{\prime\prime} - \xi^{\prime\prime} \geq 0$. 
      \item $\overline{\mathsf{trc}}(\xi,k_1,k_2) = -\mathsf{cut}(\xi,k_1,k_2) - 1\ \text{mod}\ 2^{\ell-k_1-k_2}$ if $R^{\prime\prime} - \xi^{\prime\prime} < 0$. 
    \end{enumerate}
\end{itemize}
\end{lemma}
\begin{lemma} \label{lmm:pattern3}
  \
  \begin{itemize} [leftmargin=*]
    \item If $x$ is positive, then $\overline{\mathsf{trc}}(\xi,\lambda-1) = 1$ or $\overline{\mathsf{trc}}(\xi,\lambda) = 1$. 
    \item If $x$ is negative, then $\overline{\mathsf{trc}}(\xi,\lambda-1) = 2^\ell - 1$ or $\overline{\mathsf{trc}}(\xi,\lambda) = 2^\ell - 1$. 
  \end{itemize}
\end{lemma}
\begin{IEEEproof}
If x is positive, by Lemma~\ref{lmm:pattern1}, $\overline{\mathsf{trc}}(\xi,\lambda-1) = \mathsf{cut}(\xi,\lambda-1) + \mathsf{bit} = 1 + \mathsf{bit}$.
If $\mathsf{bit} = 0$, then $\overline{\mathsf{trc}}(\xi,\lambda-1) = 1$. If $\mathsf{bit} = 1$, then $R_{\lambda-1}^{\prime\prime}+\xi_{\lambda-1}^{\prime\prime} \geq 2^{\lambda -1}$ and:
\begin{align*}
  &\xi + R = (R_{\lambda}^{\prime} + \xi_{\lambda}^{\prime})\cdot2^\ell + R_{\lambda}^{\prime\prime} + \xi_{\lambda}^{\prime\prime}\\
  =&(R_{\lambda}^{\prime} + \xi_{\lambda}^{\prime})\cdot2^\ell + R_{\lambda-1}^{\prime}\cdot2^{\lambda-1} \\
  &+ R_{\lambda-1}^{\prime\prime} + \xi_{\lambda-1}^{\prime}\cdot2^{\lambda-1} + \xi_{\lambda-1}^{\prime\prime}
\end{align*}
hence, $R_{\lambda}^{\prime\prime} + \xi_{\lambda}^{\prime\prime} \geq 2^\lambda$ and by Lemma~\ref{lmm:pattern1}, $\overline{\mathsf{trc}}(\xi,\lambda) = 0 + 1 = 1$. If x is negative,by Lemma~\ref{lmm:pattern1}, $\overline{\mathsf{trc}}(2^\ell - \xi,\lambda-1) = -\mathsf{cut}(\xi,\lambda-1) - \mathsf{bit} = 2^\ell - 1$.
If $\mathsf{bit} = 0$, then $\overline{\mathsf{trc}}(2^\ell - \xi,\lambda-1) = q - 1$. If $\mathsf{bit} = 1$, then $R_{\lambda-1}^{\prime\prime}-\xi_{\lambda-1}^{\prime\prime} < 0$ and:
\begin{align*}
  &R - \xi = (R_{\lambda}^{\prime} - \xi_{\lambda}^{\prime})\cdot2^\ell + (R_{\lambda}^{\prime\prime} - \xi_{\lambda}^{\prime\prime})\\
  =&(R_{\lambda}^{\prime} - \xi_{\lambda}^{\prime})\cdot2^\ell + R_{\lambda-1}^{\prime}\cdot2^{\lambda-1} \\
  &+ R_{\lambda-1}^{\prime\prime} - \xi_{\lambda-1}^{\prime}\cdot2^{\lambda-1} - \xi_{\lambda-1}^{\prime\prime}
\end{align*}
hence, $R_{\lambda}^{\prime\prime} - \xi_{\lambda}^{\prime\prime} < 0$ and by Lemma~\ref{lmm:pattern1}, $\overline{\mathsf{trc}}(2^\ell - \xi,\lambda) = 0 + 1 = 1$\\
\end{IEEEproof}

\begin{algorithm}[b!]
  \textbf{Setting}: $\ell$, $\ell_x$, and $p$. $P_0$ and $P_1$ share $\mathsf{seed}_{01}$, $P_0$ and $P_2$ share $\mathsf{seed}_{02}$, and $P_1$ and $P_2$ share $\mathsf{seed}_{12}$.\\
  \textbf{Preprocessing}: 
  \begin{algorithmic}[1]
    \Statex // \textit{$P_0$, $P_1$, and $P_2$ share a preprocessed Beaver triple.}
    \State $P_0$ and $P_2$ generate $[a]_0$, $[b]_0$, and $[c]_0$ using $\mathsf{seed}_{02}$,
    \Statex and $P_1$ and $P_2$ generate $[a]_1$ and $[b]_1$ using $\mathsf{seed}_{12}$. 
    \Statex Then, $P_2$ computes $[c]_1 = ([a]_0 + [a]_1)\cdot([b]_0 + [b]_1) - [c]_0\ \text{mod}\ 2^\ell$, 
    \Statex and sends $[c]_1$ to $P_1$. 
  \end{algorithmic}
  \textbf{Input}: shares of $x$\\
  \textbf{Output}: shares of $\text{ReLU}(x)$
  \begin{algorithmic}[1]
    \Statex // \textit{$P_0$,$P_1$ initialize.}
    \State $P_0$ and $P_1$ invoke Alg.~\ref{alg:drelu}, and send $[\{w_i\}]$ to $P_2$.
      \Statex // \textit{$P_2$ processes.}
      \State $P_2$ reconstructs $\{w_i\}$, and sets $\text{DReLU}(x)^\prime:=1$ if there
      \Statex exists 0 in $\{w_i\}$, otherwise sets $\text{DReLU}(x)^\prime:=0$.
      \State $P_2$ responds $e:=\text{DReLU}(x)^\prime-b\ \text{mod}\ 2^\ell$ to $P_0$ and $P_1$,
      \Statex and sends $[c]_1$ to $P_1$.
      \Statex // \textit{$P_0$ and $P_1$ finalize.}
      \State $P_0$ and $P_1$ reconstruct $[d]:=[x]-[a]\ \text{mod}\ 2^{\ell}$.
      \State $P_0$ and $P_1$ compute $[\text{ReLU}(x)]$
      \Statex $=[x\cdot \text{DReLU}(x)]=[x\cdot (\text{DReLU}(x)^{\prime} \oplus t) ]$
      \Statex $=t[x] + (1-2t)\cdot[x\cdot \text{DReLU}(x)^{\prime}]$
      \Statex $=t[x] + (1-2t)\cdot(de+d[b]+e[a]+[c])\ \text{mod}\ 2^\ell$.
    \end{algorithmic}
  \caption{UBL Bicoptor 2.0 ReLU Protocol.}
  \label{alg:relu}
\end{algorithm}
\begin{algorithm}[b!]
	\textbf{Setting}: $\ell$, $\ell_x$, and $p$. $P_0$ and $P_1$ share $\mathsf{seed}_{01}$, $P_0$ and $P_2$ share $\mathsf{seed}_{02}$, and $P_1$ and $P_2$ share $\mathsf{seed}_{12}$. $P_0$, $P_1$, and $P_2$ share $\mathsf{seed}_{012}$. $P_2$ has $\mathsf{seed}_{2}$.\\
  	\textbf{Preprocessing}: 
  	\begin{algorithmic}[1]
  		\State $P_0$ and $P_1$ generate a random bit $t$ from $\textsf{seed}_{01}$. $P_0,P_1,P_2$ generate $[\alpha]_0,[\alpha]_1,[\alpha]_2$ from $\textsf{seed}_{012}$, respectively. 
  		\Statex $P_0$ computes $[\beta]_0: = [\alpha]_0 - [\alpha]_2$, $[\beta]_1: = [\alpha]_1 - [\alpha]_0$, $[t]_0: = [\beta]_0$, and $[t]_1: = [\beta]_1 + t$.
		\Statex $P_1$ computes $[\beta]_1: = [\alpha]_1 - [\alpha]_0$, $[\beta]_2: = [\alpha]_2 - [\alpha]_1$, $[t]_1: = [\beta]_1 + t$, and $[t]_2: = [\beta]_2$. 
		\Statex $P_2$ computes $[\beta]_2: = [\alpha]_2 - [\alpha]_1$, $[\beta]_0: = [\alpha]_0 - [\alpha]_2$, $[t]_2: = [\beta]_2$, and $[t]_0: = [\beta]_0$.
        \State $P_0$,$P_1$,$P_2$ generate $[s]_1$ from $\textsf{seed}_{012}$. $P_1$ generates $[s]_2$ from $\textsf{seed}_{12}$. $P_2$ generates a random bit $s$ from $\textsf{seed}_{2}$, $[s]_2$ from $\textsf{seed}_{12}$, computes $[s]_0 = s - [s]_1 - [s]_2$, and sends $[s]_0$ to $P_0$.
        \State $P_0$,$P_1$,$P_2$ compute $[s \oplus t] = [s] + [t] - 2\cdot[s]\cdot[t]$.
   \end{algorithmic}
  \textbf{Input}: shares of $x$\\
  \textbf{Output}: shares of $\text{DReLU}(x)$ 
  \begin{algorithmic}[1]
        \Statex // \textit{$P_0$ and $P_1$ initialize.}
        \State $P_0$ sets $[x]_0+[x]_1$ as its DReLU input,
        \Statex $P_1$ sets $[x]_2$ as its DReLU input; 
        \Statex $P_0$ and $P_1$ invoke Alg.~\ref{alg:drelu}, and send $[\{w_i\}]$ to $P_2$.
        \Statex // \textit{$P_2$ processes.}
        \State $P_2$ reconstructs $\{w_i\}$, and sets $\text{DReLU}(x)^\prime:=1$ if there
      	\Statex exists 0 in $\{w_i\}$, otherwise sets $\text{DReLU}(x)^\prime:=0$.
        \State $P_2$ computes $\text{DReLU}(x)^{\prime\prime} := s \oplus \text{DReLU}(x)^\prime\ \text{mod}\ 2^\ell$.
        \State $P_2$ sends $\text{DReLU}(x)^{\prime\prime}$ to $P_0$ and $P_1$.
        \Statex // \textit{$P_0$, $P_1$, and $P_2$ finalize.}
        \State $P_0$,$P_1$,$P_2$ compute $[\text{DReLU}(x)]:=[\text{DReLU}(x)^{\prime\prime} \oplus s \oplus t] = \text{DReLU}(x)^{\prime\prime} + [s \oplus t] - 2\cdot\text{DReLU}(x)^{\prime\prime}\cdot[s \oplus t]\ \text{mod}\ 2^\ell$
      \end{algorithmic}
	\caption{RSS Bicoptor 2.0 DReLU Protocol.}
	\label{alg:drelu2}
\end{algorithm}

\begin{lemma} \label{lmm:pattern4}
  If $\lambda + 1 < \ell$, for any $\lambda^\prime > \lambda$, 
    \begin{itemize} [leftmargin=*]
      \item If $\overline{\mathsf{trc}}(\xi,\lambda^\prime-1) = 1$, then $\overline{\mathsf{trc}}(\xi,\lambda^\prime) = 1$ or $0$.
      \item If $\overline{\mathsf{trc}}(2^\ell - \xi,\lambda^\prime-1) = 2^\ell - 1$, then $\overline{\mathsf{trc}}(\xi,\lambda^\prime) = 1$ or $0$.
    \end{itemize}
\end{lemma}
\begin{IEEEproof}
  \textbf{Case 1}: For $\overline{\mathsf{trc}}(\xi,\lambda^\prime-1) = 1$, we have $\overline{\mathsf{trc}}(\xi,\lambda^\prime-1) = \mathsf{cut}(\xi,\lambda^{\prime}-1) + \mathsf{bit}_{\lambda^{\prime}-1}$. If $\mathsf{cut}(\xi,\lambda^{\prime}-1) = 1$ and $\mathsf{bit}_{\lambda^{\prime}-1} = 0$, then $\mathsf{cut}(\xi,\lambda^{\prime}) = 0$ and therefore $\overline{\mathsf{trc}}(\xi,\lambda^\prime) = \mathsf{cut}(\xi,\lambda^{\prime}) + \mathsf{bit}_{\lambda^\prime} = 0 + \mathsf{bit}_{\lambda^\prime} = 0$ or $1$. If $\mathsf{cut}(\xi,\lambda^{\prime}-1) = 0$ and $\mathsf{bit}_{\lambda^{\prime}-1} = 1$, then $\mathsf{cut}(\xi,\lambda^{\prime}) = 0$ and therefore $\overline{\mathsf{trc}}(\xi,\lambda^\prime) = \mathsf{cut}(\xi,\lambda^{\prime}) + \mathsf{bit}_{\lambda^\prime} = 0 + \mathsf{bit}_{\lambda^\prime} = 0$ or $1$. 

  \textbf{Case 2}: For $\overline{\mathsf{trc}}(2^\ell - \xi,\lambda^\prime-1) = 2^\ell - 1$, we have $\overline{\mathsf{trc}}(2^\ell - \xi,\lambda^\prime-1) = - \mathsf{cut}(\xi,\lambda^{\prime}-1) - \mathsf{bit}_{\lambda^{\prime}-1}\ \text{mod}\ 2^{\ell}$. The rest of the proof is similar to that in Case 1.
\end{IEEEproof}

\begin{lemma} \label{lmm:pattern5}
  If $\lambda + 1 < \ell$, for any $\lambda^\prime > \lambda$ we have $\overline{\mathsf{trc}}(\xi,\lambda^\prime) = 0$ if $\overline{\mathsf{trc}}(\xi,\lambda^\prime-1) = 0$.
\end{lemma}
\begin{IEEEproof}
  \textbf{Case 1}: For $x = \xi$, $\overline{\mathsf{trc}}(\xi,\lambda^\prime-1) = \mathsf{cut}(\xi,\lambda^{\prime}-1) + \mathsf{bit}_{\lambda^\prime-1}$. Since $\overline{\mathsf{trc}}(\xi,\lambda^\prime-1) = 0$ which implies $\mathsf{cut}(\xi,\lambda^{\prime}-1) = 0$ and $\mathsf{bit}_{\lambda^\prime-1} = 0$. By Lemma~\ref{lmm:pattern1}, $r_{\lambda^\prime-1}^{\prime\prime} + \xi_{\lambda^\prime-1}^{\prime\prime} < 2^{\lambda^\prime-1}$ and hence $R_{\lambda^\prime}^{\prime\prime} + \xi_{\lambda^\prime}^{\prime\prime} = R_{\lambda^\prime-1}^{\prime}\cdot2^{\lambda^\prime-1} + R_{\lambda^\prime-1}^{\prime\prime} + \xi_{\lambda^\prime-1}^{\prime}\cdot2^{\lambda^\prime-1} + \xi_{\lambda^\prime-1}^{\prime\prime} = 0\cdot2^{\lambda^\prime-1} + R_{\lambda^\prime-1}^{\prime\prime} + 0^{\prime}\cdot2^{\lambda^\prime-1} + \xi_{\lambda^\prime-1}^{\prime\prime} = R_{\lambda^\prime-1}^{\prime\prime} + \xi_{\lambda^\prime-1}^{\prime\prime} < 2^{\lambda^\prime-1}$ which finally implies $\overline{\mathsf{trc}}(\xi,\lambda^\prime) = 0 + 0 = 0$.

  \textbf{Case 2}: For $x = 2^\ell - \xi$, $\overline{\mathsf{trc}}(2^\ell - \xi,\lambda^\prime-1) = - \mathsf{cut}(\xi,\lambda^{\prime}-1) - \mathsf{bit}_{\lambda^\prime-1}\ \text{mod}\ 2^{\ell}$.The rest of the proof is similar to that in Case 1. 
\end{IEEEproof}

\subsection{How to construct the UBL Bicoptor 2.0 ReLU protocol} \label{app:relu}
Alg.~\ref{alg:relu} describes the UBL Bicoptor 2.0 ReLU protocol constructed based on the UBL Bicoptor 2.0 DReLU protocol (Alg.~\ref{alg:drelu}), where the multiplication is accomplished via preprocessed triples. For more information on how to generate triples during the online phase in ReLU protocols, please refer to Bicoptor~\cite{ZWC+-23}.

\subsection{How to construct the RSS Bicoptor 2.0 DReLU and ReLU protocols} \label{app:rss}
Alg.~\ref{alg:drelu2} describes the RSS Bicoptor 2.0 DReLU protocol, where each participant holds 2-out-of-3 shares of the input $x$. The RSS Bicoptor 2.0 ReLU protocol can be constructed by simply invoking the secret multiplication protocol for $[x\cdot \text{DReLU}(x)]$.


\subsection{The Communication Amount and Round of Previous DReLU Protocols} \label{app:calc}
\noindent \textbf{Falcon.} In Falcon's DReLU protocol, the theoretical communication cost of the protocol is determined by the combination of the Wrap function $\Pi_\mathsf{WA}$~\cite[Alg.~2]{WTB+-21} and some local computations. Consequently, all communication in the DReLU protocol is derived from $\Pi_\mathsf{WA}$~\cite[Alg.~2]{WTB+-21}. The communication rounds required by the Wrap function are composed of two parts: one round of communication required for the reconstruct operation (step 2) and the communication rounds associated with the Private Compare function $\Pi_\mathsf{PC}$~\cite[Alg.~1]{WTB+-21}. The $\Pi_\mathsf{PC}$~\cite[Alg.~1]{WTB+-21} involves $\log\ell + 2$ rounds of communication, with the multiplication in step 2 being computed in parallel and completed within one round of communication. The remaining $\log\ell + 1$ rounds of communication are attributed to the $\Pi_\mathsf{PC}$~\cite[Alg.~1]{WTB+-21} in step 6. Lastly, the output $[b]_2$ of the DReLU protocol is transformed to $[b]_{2^\ell}$, which adds one more round of communication. Altogether, the DReLU protocol requires $\log\ell + 4$ rounds of communication. The total communication amount is $17\ell$, where $16\ell$ comes from step 2 and 6 in $\Pi_\mathsf{PC}$~\cite[Alg.~1]{WTB+-21}, and $\ell$ comes from step 2 in $\Pi_\mathsf{WA}$~\cite[Alg.~2]{WTB+-21}.

\noindent \textbf{Edabits.} According to the description in section 5.2 of the Edabits~\cite{EGK+-20b}, we can use truncation protocols to construct Integer Comparison. The paper introduces four truncation protocols, and we mainly focus on the communication cost of using the \textsf{LogicalRightShift}~\cite[Fig.~9]{EGK+-20b} to construct the comparison.

The communication rounds are calculated as follows:
\begin{itemize} [leftmargin=*]
  \item Step 1(a) and 1(c) each require one round of communication.
  \item Step 1(b), using the LT (Less Than) protocol, requires $\log\ell$ rounds of communication.
  \item Steps 2(b) and 2(d) each need one round of communication.
  \item Step 2(c) does not require communication because $m = \ell - 1$.
\end{itemize}

In total, the communication cost is $4 + \log\ell$ rounds. The total communication amount is $4\ell-2$, where $2(\ell-1)$ comes from step 1(b), as the LT protocol consumes $\ell - 1$ boolean triples, and each triple requires $2$ bits of communication. Additionally, $2\ell$ of communication is from step 1(a) and (2b). We can ignore the communication amount for step 1(c) and 2(d).

\noindent \textbf{One-pass Dominating Communication Cost with Key-Bits Optimization (lower section of Tab.~\ref{tab:compare})} In the computation of communication costs for Falcon and Edabits, the term $\ell - 24$ appears due to $\ell = 64$ and the original precision being represented as $\ell_x=5+26$, indicating 5 bits for the integer part and 26 bits for the fractional part. After applying the key bits optimization, we represent the precision as $\ell_x=5+2$, effectively ignoring the last $24$ bits of the fractional part. Consequently, in the calculation of communication costs, we use $\ell - 24$ to reflect this reduced fractional precision.

\end{document}